\documentclass[twocolumn,aps,floatfix,citeautoscript,nofootinbib,superscriptaddress]{revtex4}
\usepackage{amsbsy}
\usepackage{latexsym,epsfig,graphicx}
\usepackage{dcolumn}
\usepackage{graphicx}
\usepackage{subfigure}
\usepackage{comment}
\usepackage{color}
\usepackage{bm}
\usepackage{mathrsfs}
\usepackage{amsfonts}
\usepackage{amsmath}
\usepackage{color}
\usepackage{amssymb}
\usepackage{xspace}
\usepackage{epstopdf}
\usepackage{tabularx}
\usepackage{longtable}
\usepackage[colorlinks=true, letterpaper=true, pdfstartview=FitV, linkcolor=blue, citecolor=blue, urlcolor=blue]{hyperref}
\usepackage[normalem]{ulem}

\pdfoutput=1
\begin{document}

\title{Magnetic flux induced higher-order topological superconductivity}
\author{Jinpeng Xiao}
    \altaffiliation{Corresponding author: xiaojinpeng2018@163.com}
	\affiliation{
		School of Mathematics and Physics, Key Laboratory of Energy Conversion Optoelectronic Functional Materials of Jiangxi Education Institutes, Jinggangshan University, Ji'an 343009, China}

	\author{Qianglin Hu}
    \affiliation{
		Department of Physics and Electronics Engineering. Tongren University, Tong Ren 554300, China}
	\author{Zuodong Yu}
    \affiliation{
		School of Information and Electronic Engineering, Zhejiang Gongshang University, Hangzhou, 310018, China}

 \author{Weipeng Chen}
    \altaffiliation{Corresponding author: wpchen2023@163.com}
    \affiliation{
		Department of Physics and Siyuan Laboratory, Jinan University, Guangzhou 510632, China}

	\author{Xiaobing Luo}
   \altaffiliation{Corresponding author: xiaobingluo2013@aliyun.com}
   \affiliation{Zhejiang Key Laboratory of Quantum State Control and Optical Field Manipulation, Department of Physics, Zhejiang Sci-Tech University, Hangzhou, 310018,  China}

\begin{abstract}

Higher-order topological superconductivity typically depends on spin-orbit interaction, and often necessitates well designed sample structures, nodal superconducting pairings or complex magnetic order. In this work, we propose a model that incorporates a Zeeman field, antiferromagnetic order, and $s$-wave superconducting pairing, all without the need for spin-orbit interaction. In a two-dimensional system, we realize a second-order topological superconductor by utilizing a staggered flux, provided that the Zeeman field is oriented perpendicular to the magnetic order moments. In three-dimensional systems, we achieve second- and third-order topological superconductors in theory, through stacking the two-dimensional second-order topological superconductor.
\end{abstract}

\maketitle

\section{Introduction}
Superconducting surfaces coated with magnetic atoms can induce Yu-Shiba-Rusinov (YSR) states within the superconducting energy gap\cite{Yu1965,Shiba1968,Rusinov1969}, facilitating the formation of various types of topological superconductivity, which has garnered significant attention\cite{1,2,3,4,75,5,6,7,8,9,10,11,12,13,14,15,16,17,18}. With advancements in experimental techniques, scanning tunneling microscopy (STM) not only enables the manipulation of individual magnetic atoms on superconducting surfaces but also provides greater insights into YSR states\cite{19,20,21,22,23}. Recently, a variety of anisotropic YSR states have been observed on $s$-wave superconductor surfaces, closely linked to the orbital characteristics of valence electrons in magnetic atoms\cite{37,24,25,85,26,27,28,29,30,31,32,87,88}. Moreover, sometimes these YSR states can extend over several tens of nanometers, spanning scales of hundreds of atoms\cite{37,32}, thus offering a promising platform for the artificial realization of magnetic topological superconductors (TSCs).

Recently, a novel topological superconducting phase known as higher-order topological superconductivity has been proposed\cite{39,40,41,42,43,44,442,45,46,47,48,49,50,51,52,53,54,55,56,562,57,58,59,60,61,62,63,64,65,66,67,70}. Its most significant distinction from traditional TSCs is the emergence of Majorana zero-energy boundary states, which manifest as Majorana corner modes (MCMs) and Majorana hinge modes (MHMs). For example, in a $d$ dimensional $n$th-order TSC, the Majorana zero modes emerge on ($d-n$) (where $n>1$) dimensional boundaries rather than on the conventional $(d-1)$ dimensional boundaries. This phenomenon can be elucidated by low-energy edge theory\cite{43}. In a $d$ dimensional $n$th-order TSC, the energy gap functions of the effective edge Hamiltonians at two adjacent ($d-n+1$) dimensional boundaries are non-zero yet possess opposite signs, leading to the formation of domain walls between them that bind zero-energy bound states\cite{69}, corresponding to Majorana corner states or hinge states.
\begin{figure}[t]
\centering\includegraphics[width=0.3\textwidth]{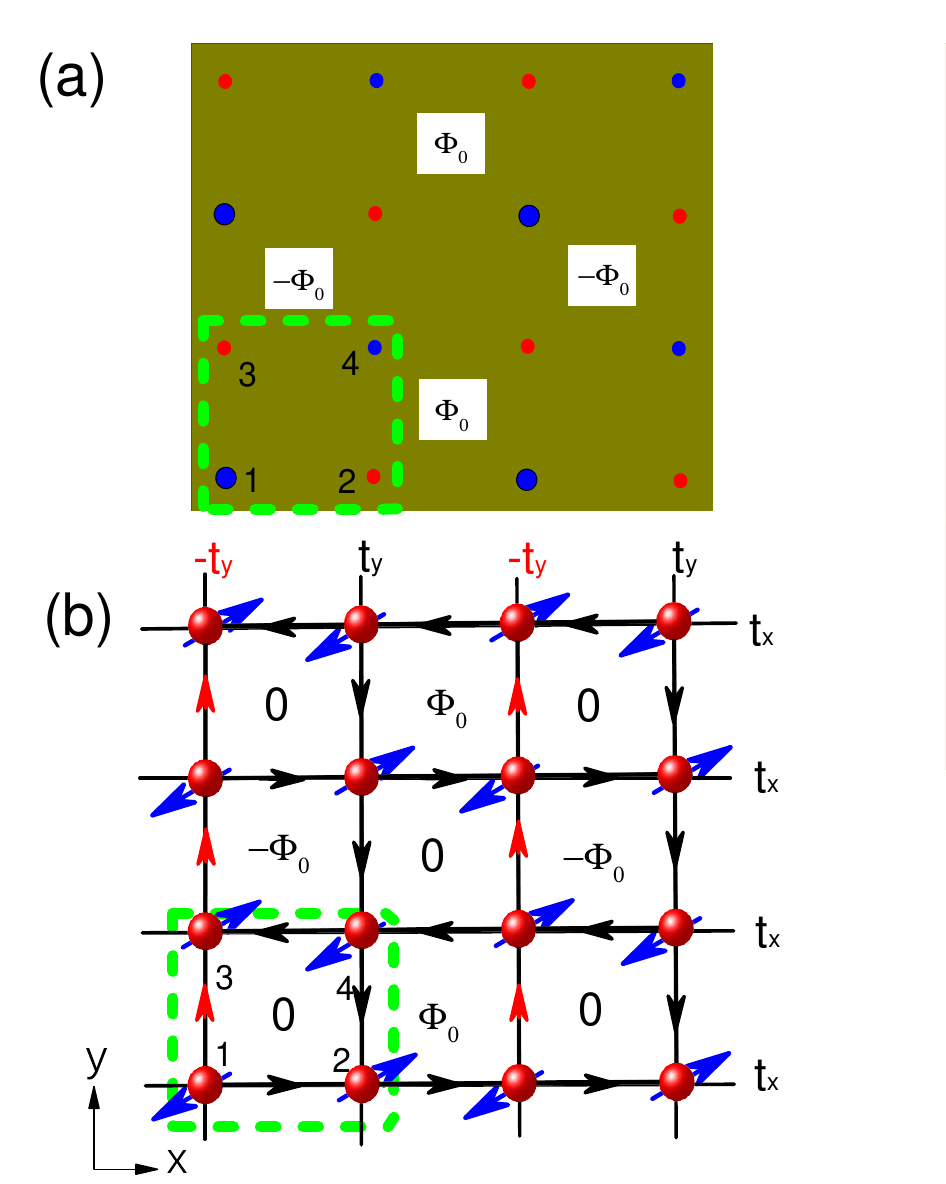}
\caption{(a) The schematic of two-dimensional square arrays of magnetic atoms deposited on an $s$-wave superconductor. The blue and red dots symbolize opposite spins, while the larger blue dots represent different atoms due to their distinct spatial wave functions. The periodically arranged holes contain staggered quantum flux $\Phi_{0}$ through them. (b) The schematic of the effective tight-binding lattice model. The arrows on the bonds indicate the hopping direction with a positive phase factor, while the colors denote different signs of hopping amplitudes. Each unit cell comprises four atoms, as indicated by the green dashed-line box.
}
\label{Fig1}
\end{figure}
Numerous studies have reported the achievement of higher-order topological superconductors (HOTSCs) through various approaches, including the utilization of Josephson junctions\cite{39,40,41,42}, nodal superconducting pairing\cite{43,44,442,45,46,47,48,49,50,51,52}, magnetic atomic configurations\cite{17,53,54,55,70,562}, and the proximity effect between topological insulators and superconductors\cite{442,56,562,57,58,59,60,61,62,41,42}. However, in these investigations, the formation of domain walls between adjacent effective edge Hamiltonians is dependent on the spin-orbit interaction (SOI). On the other hand, proposals utilizing magnetic atomic order mostly require nodal superconducting pairings or complex magnetic structures.

In this paper, we propose a method to realize HOTSCs in conventional $s$-wave superconductors without the need for SOI and complex magnetic configurations. This approach employs staggered magnetic flux and relies solely on the YSR states arising from antiferromagnetic (AFM) order, along with an external magnetic field that is perpendicular to the magnetic moments.

\section{Model}

We consider magnetic atoms deposited antiferromagnetically on a two-dimensional (2D) $s$-wave superconductor. The 2D superconductor is periodically hollowed out, and staggered quantum magnetic flux is arranged as illustrated in Fig. \ref{Fig1}(a).
The red and blue dots represent magnetic atoms with antiparallel spins located on the surface of the superconductor. Notably, atom 1 differs from the other three atoms; its YSR state exhibits anisotropy in the $x$- and $y$-directions, for example $d_{x^{2}-y^{2}}$-like\cite{24,25,27}, resulting in different signs of the overlap integral of wave functions between neighboring atoms in the $x$- and $y$-directions. Here, we employ a classical spin approximation for the magnetic atoms, wherein they act as on-site magnetic fields.
Utilizing STM techniques, magnetic adatoms can be deposited onto the superconducting substrate with atomic precision. Under appropriate doping conditions, the array of magnetic adatoms self-organizes into an AFM configuration, facilitated by the Ruderman-Kittel-Kasuya-Yosida (RKKY) interaction within a finite-size system\cite{75,5,95}.
The schematic representation of the effective tight-binding model is presented in Fig. \ref{Fig1}(b), which consists of chains of 1D TSCs\cite{18}. The mean-field BCS model Hamiltonian can be expressed as follows:
\begin{eqnarray}\label{eq1}
H&=&\sum_{<jl>\alpha}t_{jl}e^{-i\frac{\phi_{jl}}{2}}c_{j}^{\dag}c_{l}+Jc_{j}^{\dag}\mathbf{S}_{j}\cdot\mathbf{\sigma}c_{j}+\mu c_{j}^{\dag}c_{j}\nonumber\\
&+&V c_{j}^{\dag}\sigma_z c_{j}+\Delta c_{j\uparrow}^{\dag}c_{j\downarrow}^{\dag}+\Delta c_{j\downarrow}c_{j\uparrow}.
\end{eqnarray}
$c_{j}^{\dag}=(c_{j\uparrow}^{\dag},c_{j\downarrow}^{\dag})$ refers to the electron creation operators at site $j$. The first term represents the hopping term, where $t_{jl}$ is the hopping amplitude of itinerant electrons between nearest neighboring magnetic atoms, and the phase factor is given by $\phi_{jl}/2=\phi_{j}/2-\phi_{l}/2$. $\phi_{j}/2$ denotes the phase factor of the single-particle wave function. The quantization of magnetic flux mandates that the phase $\phi_{j}$ changes by $2\pi n$ around the four magnetic atoms in one unit cell. Assuming $n=1$, we have $\phi_{jl}=\pi/2$, resulting in a hopping phase factor of $\pm\pi/4$. The second term accounts for the in-plane antiferromagnetically arranged magnetic atoms, described by $\mathbf{S}_{j}=S(\cos\varphi,\sin\varphi,0)$, where $\varphi$ is a random angle. $J$ stands for the exchange coupling constant between the itinerant electrons and the on-site magnetic moment, while $\boldsymbol{\sigma}=(\sigma_x,\sigma_y,\sigma_z)$ represents the vector of Pauli spin matrices. The third and fourth terms correspond to the chemical potential and the external Zeeman field, respectively. A Zeeman field generally distorts the AFM order established via the RKKY mechanism\cite{6,18}; here, we assume the Zeeman field is sufficiently weak so as not to completely destroy the AFM order. The final two terms represent the $s$-wave pairing interactions.

Our model employs an idealized scenario, assuming a high-quality superconducting substrate and neglecting boundary disorder near the holes. While this simplification allows for a tractable theoretical analysis, it omits the effects of potential imperfections and boundary disorder on the Shiba impurity lattice.

\section{Topological phase diagram and Majorana corner modes}
\begin{figure}[t]
\centering\includegraphics[width=0.45\textwidth]{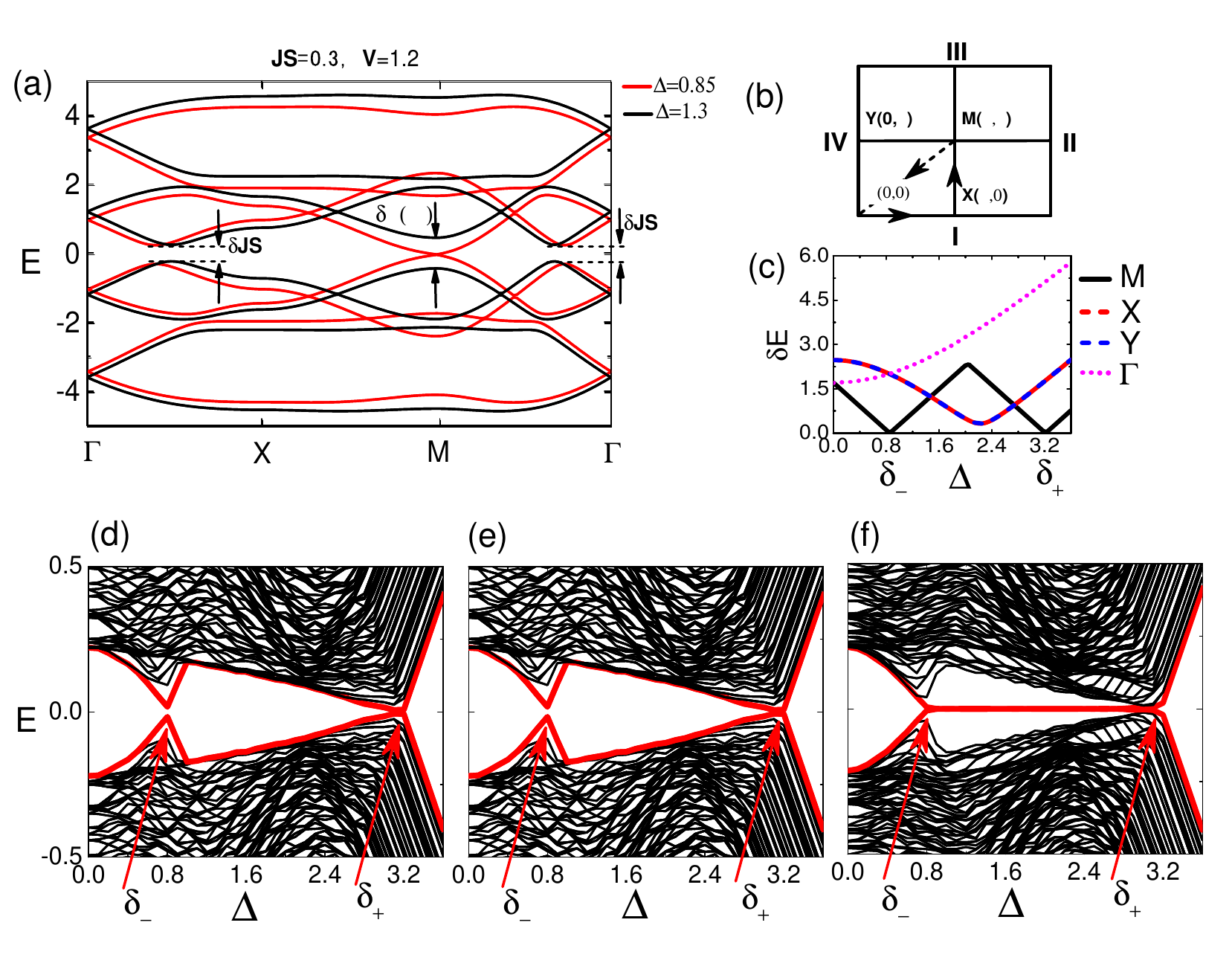}
\caption{The energy bands and open boundary spectra of Hamiltonian (\ref{eq2}). (a) Shows the dispersions along the path depicted in (b), which represents the 2D Brillouin Zone (BZ). (c) The gaps at the four HSPs in the BZ as a function of $\Delta$ with the energies at $X$ and $Y$ exhibiting two-fold degeneracy. (d) and (e) display the spectra of a system with 24$\times$24 sites under OBCs applied in the $x$- or $y$-directions, while (f) presents the case with OBCs imposed in both directions.}
\label{Fig2}
\end{figure}
The Hamiltonian can be expressed in momentum space as $H=\sum_{\mathbf{k}}\Psi_{\mathbf{k}}^{\dag}H(\mathbf{\mathbf{k}})\Psi_{\mathbf{k}}$, where the basis spinor is $\Psi_{\mathbf{k}}=[f_{1},f_{2},f_{3},f_{4}]^{T}$ with $f_{\delta}=[c_{\delta \mathbf{k}\uparrow},c_{ \delta \mathbf{k}\downarrow},c_{\delta -\mathbf{k}\downarrow}^{\dag},-c_{\delta -\mathbf{k}\uparrow}^{\dag}]$. Here, $c_{ \delta \mathbf{k}\alpha}=\frac{1}{\sqrt{N}}\sum_{j}e^{-i\mathbf{k}\cdot \mathbf{R}_{j\delta}}c_{ \delta j\alpha}$, where $\alpha$ denotes the spins labels, $\delta$ represents the sublattice sites, and $N$ denotes the number of unit cells. Then we have
\begin{eqnarray}\label{eq2}
H(\mathbf{k})=H_{x}+H_{y}+H_{on},
\end{eqnarray}
with
\begin{eqnarray}\label{eq3}
H_{x}&=&\xi_{x}[(1+\cos k_{x})s_{x}\tau_z-\sin k_{x}s_{y}\tau_z]\nonumber\\
&+&\eta_{x}[(1-\cos k_{x})s_{y}\Gamma_z-\sin k_{x}s_{x}\Gamma_z],\nonumber\\
H_{y}&=&\xi_{y}[(1+\cos k_{y})s_{z}\Gamma_x\tau_z-\sin k_{y}s_{z}\Gamma_y\tau_z]\nonumber\\
&+&\eta_{y}[(1-\cos k_{y})\Gamma_y-\sin k_{y}\Gamma_x],\nonumber\\
H_{on}&=&\Delta\tau_x+(JS_{x}\sigma_x+JS_{y}\sigma_y)s_{z}\Gamma_z+V\sigma_z.
\end{eqnarray}
Here $\xi_{x(y)}=t_{x(y)}\cos \phi$, $\eta_{x(y)}=t_{x(y)}\sin \phi$, $S_{x}=S\cos\varphi$ and $S_{y}=S\sin\varphi$. $t_{x(y)}$ represents the hopping amplitude along the $x(y)$-direction. $\phi=\pi/4$. The Pauli matrices $\tau_{i}$, $\sigma_{i}$, $s_{i}$ and $\Gamma_{i}$ operate in the particle-hole space, spin space, and the $x$ and $y$ sublattice spaces, respectively. For simplicity, in this work we consider $\mu=0$, $t_{x}=t_{y}=t=1.0$ and all energies expressed in units of $t$. In the normal states, the anti-commutation relationships along the terms $H_x$, $H_y$, $(JS_{x}\sigma_x+JS_{y}\sigma_y)s_{z}\Gamma_z$ ensure that the system remains gapped. When considering the superconducting pairing, at point $M$ (see Fig. \ref{Fig2}(b)), the term $\Delta\tau_x$ commutes with all other terms, indicating a potential gap closing at this point. We present the numerically calculated energy dispersions in Fig. \ref{Fig2}(a) under two different values of $\Delta$.
The gap at point $M$ closes and reopens as the strength of $\Delta$ increases, as illustrated by the red and black curves.

The eigenvalues at the four high-symmetric points (HSPs) in the 2D BZ can be calculated analytically. We provide the eigenvalues at point $M$ as follows (The eigenvalues of other points can be found in Appendix \ref{A1}):
\begin{eqnarray}\label{eq4}
\epsilon(M)=\pm\Delta\pm\delta_{\pm},
\end{eqnarray}
where
$\delta_{\pm}=\sqrt{(2\sqrt{\eta_{x}^{2}+\eta_{y}^{2}}\pm V)^{2}+J^{2}S^{2}}$.
It is evident that the gap at point $M$ closes when $\Delta=\delta_{\pm}$.
In contrast, the other bands at points $X$, $Y$ and $\Gamma$ are consistently gapped, as shown in the numerical results presented in \ref{Fig2}(c), which are consistent with the analytical derivations.

\begin{figure}[t]
\centering\includegraphics[width=0.48\textwidth]{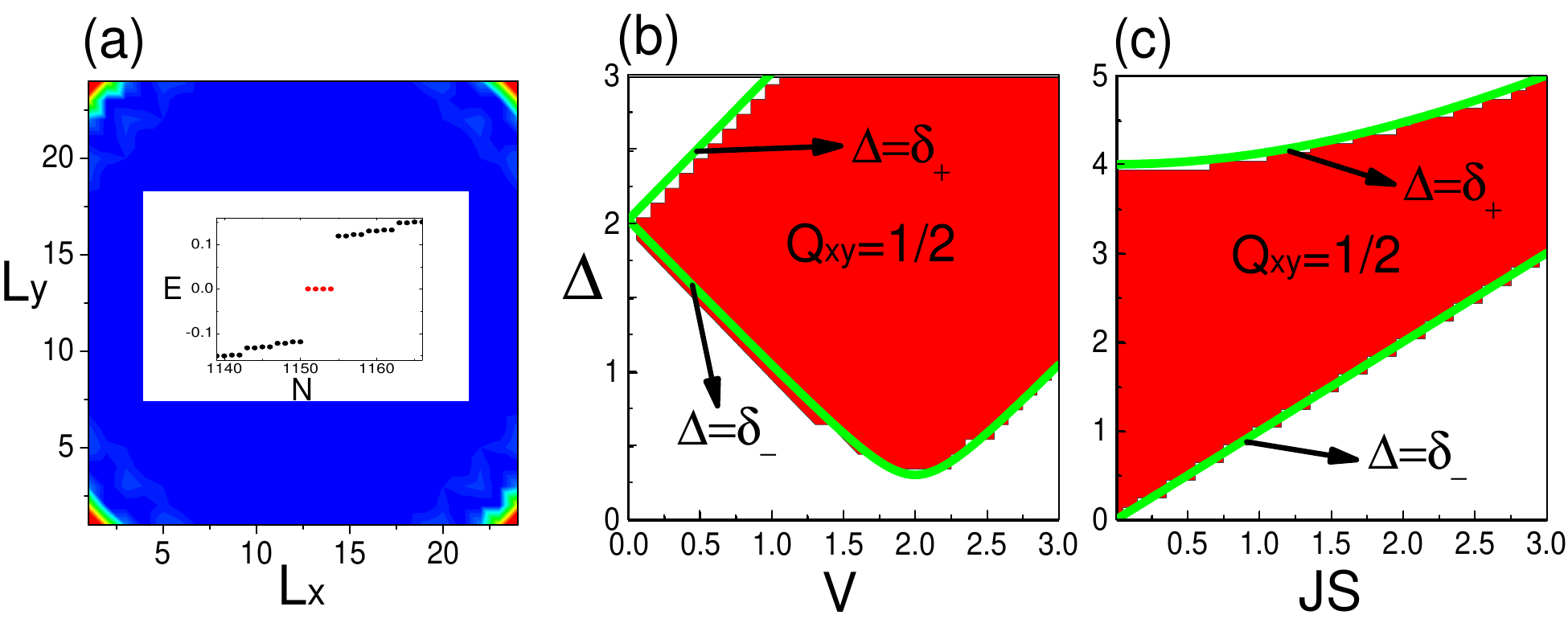}
\caption{(a) The density distributions of the four MCMs (red dots in inset) in the system comprising 24$\times$24 sites under simultaneous OBCs applied in both $x$- and $y$-directions, with $\Delta=2.0$, $JS=0.3$, $V=1.2$. (b)(c) illustrate the second order topological phase diagrams, where the red regions indicate $Q_{xy}=1/2$. The green solid lines represent the gap-closing conditions calculated by Eq. (\ref{eq4}), with $JS=0.3$ in panel (b) and $V=2.0$ in panel (c).
}
\label{Fig3}
\end{figure}

Furthermore, in Fig. \ref{Fig2}(d)[(e)], we display the spectra obtained by imposing open boundary conditions (OBCs) in the $x(y)$-direction, while applying periodic boundary conditions (PBCs) in the $y(x)$-direction. Although gap closing occurs within the spectrum of the ribbon geometry, coinciding with the bulk gap, as shown in Fig. \ref{Fig2}(c), no zero modes are present. In Fig. \ref{Fig2}(f), we illustrate the spectrum when OBCs are applied in both directions, accompanied by the emergence of zero modes. These results indicates the absence of a first-order topological phase transition, while a second-order topological phase transition may exist. In Fig. \ref{Fig3}(a), we present the density distributions of the zero-energy modes, which localize at the four corners, thereby confirming the presence of the second-order topological phase.

A 2D second-order topological superconductor (SOTSC) is characterized by a quantized bulk quadrupole moment $Q_{xy}$\cite{63,64}. In superconductors, this expression can be simplified as follows\cite{65,66,67},
\begin{eqnarray}\label{eq5}
Q_{xy}=\frac{1}{2\pi}Im\{\ln[\mathrm{Det}(U^{\dag}QU)\sqrt{\mathrm{Det}(Q^{\dag})}]\},
\end{eqnarray}
where $U$ is a $16N\times 8N$ matrix constructed by column-wise packing all the occupied eigenstates. The operator $Q=\exp[i2\pi \hat{q}_{xy}]$, with $\hat{q}_{xy}=\hat{x}\hat{y}/(L_x L_{y})$ representing the microscopic quadrupole operator, where $\hat{x}(\hat{y})$ is the position operator and $L_x(y)$ the corresponding size along $x(y)$-direction. The particle-hole symmetry ensures that $Q_{xy}$ is quantized to 1/2 or 0\cite{67}.

It is noted that there are some intrinsic issues associated with calculating the quadrupole moment by Eq. (\ref{eq5})\cite{89}. $Q_{xy}$ is not arbitrary translation invariant, therefore, not correctly evaluate the physical quadrupole moment, and the improper choice of the origin point may lead to zero values even when the system is in SOTSC phase. However, the quadrupole moment can still serve as an indicator for second-order topology of quadrupole insulators and SOTSCs\cite{90,91,92,93,94}. In the following calculation, we choose the origin point $(x,y)=(1,1)$, as well as even values for $L_{x(y)}$, therefore partly avoid the difficulties proposed in Ref. \cite{89}. Furthermore, the closing and reopening of the energy gap, along with the presence of zero-energy corner states, indicate a second-order topological phase transition and can be also used to prove the rationality of $Q_{xy}$ as the topological invariant.

By numerically solving the Hamiltonian (\ref{eq1}), we calculate the quadrupole moment $Q_{xy}$ across the parameter space, as illustrated in Figs. \ref{Fig3}(b) and (c) through color representation. The boundaries of the nontrivial phase correspond to the gap-closing conditions, marked by the green solid curves. Note that both the bulk energy gap and boundary spectrum gap close under the same condition. This indicates that the bulk quadrupole moment signifies a topological phase transition that is not merely a boundary phenomenon, but rather a transition involving both bulk and boundary physics. In the next section, we will present the phase transition and appearance of MCMs from the perspective of the low energy edge Hamiltonian.

\section{Low energy edge theory}\label{sec4}
The low energy edge theory is an useful tool for understanding the emergence of MCMs in HOTSCs. Since the band gap of Hamiltonian (\ref{eq2}) closes only at point $M$, we expand $H_x$ and $H_y$ around $M$ to the second order,
\begin{eqnarray}\label{eq6}
H_{x}&=&\frac{1}{2}(\xi_{x}s_{x}\tau_z-\eta_{x} s_{y}\Gamma_z)q_{x}^2+2\eta_{x}s_{y}\Gamma_z\nonumber\\
&+&(\eta_{x}s_{x}\Gamma_{z}+\xi_{x}s_{y}\tau_z)q_{x},\nonumber\\
H_{y}&=&\frac{1}{2}(\xi_{y}s_{z}\Gamma_x\tau_z-\eta_{y} \Gamma_y)q_{y}^2+2\eta_{y}\Gamma_y\nonumber\\
&+&(\eta_{y}\Gamma_{x}+\xi_{y}s_{z}\Gamma_y\tau_z)q_{y},
\end{eqnarray}
where $q_{x(y)}=k_{x(y)}-\pi$.
We label the four boundaries with I$-$IV, as shown in Fig. \ref{Fig2}(b). Taking edge-I as an example, we apply PBCs along the $x$-direction and OBCs along the $y$-direction, substituting $q_{y}$ with $-i\partial_{y}$ and neglecting the $q_{x}^2$ term. The Hamiltonian can be decomposed into two parts $H=H_{0}(-i\partial_{y})+H_{p}$ with

\begin{eqnarray}\label{eq7}
H_{0}&=&-\frac{1}{2}(\xi_{y}s_{z}\Gamma_x\tau_z-\eta_{y} \Gamma_y)\partial_{y}^2-i(\eta_{y}\Gamma_{x}+\xi_{y}s_{z}\Gamma_y\tau_z)\partial_{y},\nonumber\\
H_{p}&=&(\eta_{x}s_{x}\Gamma_{z}+\xi_{x}s_{y}\tau_z)q_{x}+2\eta_{y}\Gamma_y+2\eta_{x}s_{y}\Gamma_z \nonumber\\
&+&\Delta\tau_x+JS_{x}\sigma_xs_{z}\Gamma_z+JS_{y}\sigma_ys_{z}\Gamma_z+V\sigma_z.
\end{eqnarray}
We solve the equation $H_{0}\psi_{\alpha}=E_{\alpha}\psi_{\alpha}(y)$ with $E_{\alpha}=0$. By projecting $H_{p}$ onto the eigenstates $\psi_{\alpha}$, one can obtain the gapped 1D edge Hamiltonian. Following this procedure allows for the acquisition of all four edge Hamiltonians. A detailed derivation process is presented in Appendix \ref{B1}. As the edge coordinates $l$ are defined in a counterclockwise manner, the edge Hamiltonian can be compactly written as
\begin{eqnarray}\label{eq8}
H_l&=&-iA(l)s_z\tau_z\sigma_x\partial_{l}+\Delta(l)\sigma_z+H_{M},
\end{eqnarray}
where $l$=\{I$-$IV\}, $A(l)=\{\xi_{x},\xi_{y},\xi_{x},\xi_{y}\}$, $\Delta(l)=\{\Delta,-\Delta,\Delta,-\Delta\}$, $H_{M}=2\eta(l)s_z\Gamma_z\tau_z+2\eta'(l)\Gamma_y-(JS_{x}\tau_x+JS_{y}\tau_y)\Gamma_z+V\tau_z$ with $\eta(l)=\{\eta_{x},\eta_{y},\eta_{x},\eta_{y}\}$ and $\eta'(l)=\{\eta_{y},-\eta_{x},\eta_{y},-\eta_{x}\}$.
Here the Pauli matrices $s_z$, $\Gamma_{i}$, $\tau_{i}$ and $\sigma_{i}$ are defined within the space of the sixteen zero-energy states $\psi_{\alpha}$.

The edge energy spectrum of Eq. (\ref{eq8}) with $q_{l}=0$ reads $\pm\Delta\pm\delta_{\pm}$, consistent with Eq. (\ref{eq4}). The gap closes at the boundaries of the topological phase, as calculated using Eq. (\ref{eq5}), further confirming the validity of $Q_{xy}$ as the second-order topological invariant.
Without loss of generality, we assume $\delta_{+}>\Delta>\delta_{-}$. The gaps at edges-I and -III are $2(\Delta-\delta_{-})$, while at edge-II and -IV are $-2(\Delta-\delta_{-})$. The signs of the mass term are opposite at adjacent boundaries, resulting in the formation of a mass domain wall at each corner. Consequently, each corner traps one MCM, as illustrated in Fig. \ref{Fig3}(a).

\section{Hinge modes and third-order corner modes}
\begin{figure}[t]
\centering\includegraphics[width=0.45\textwidth]{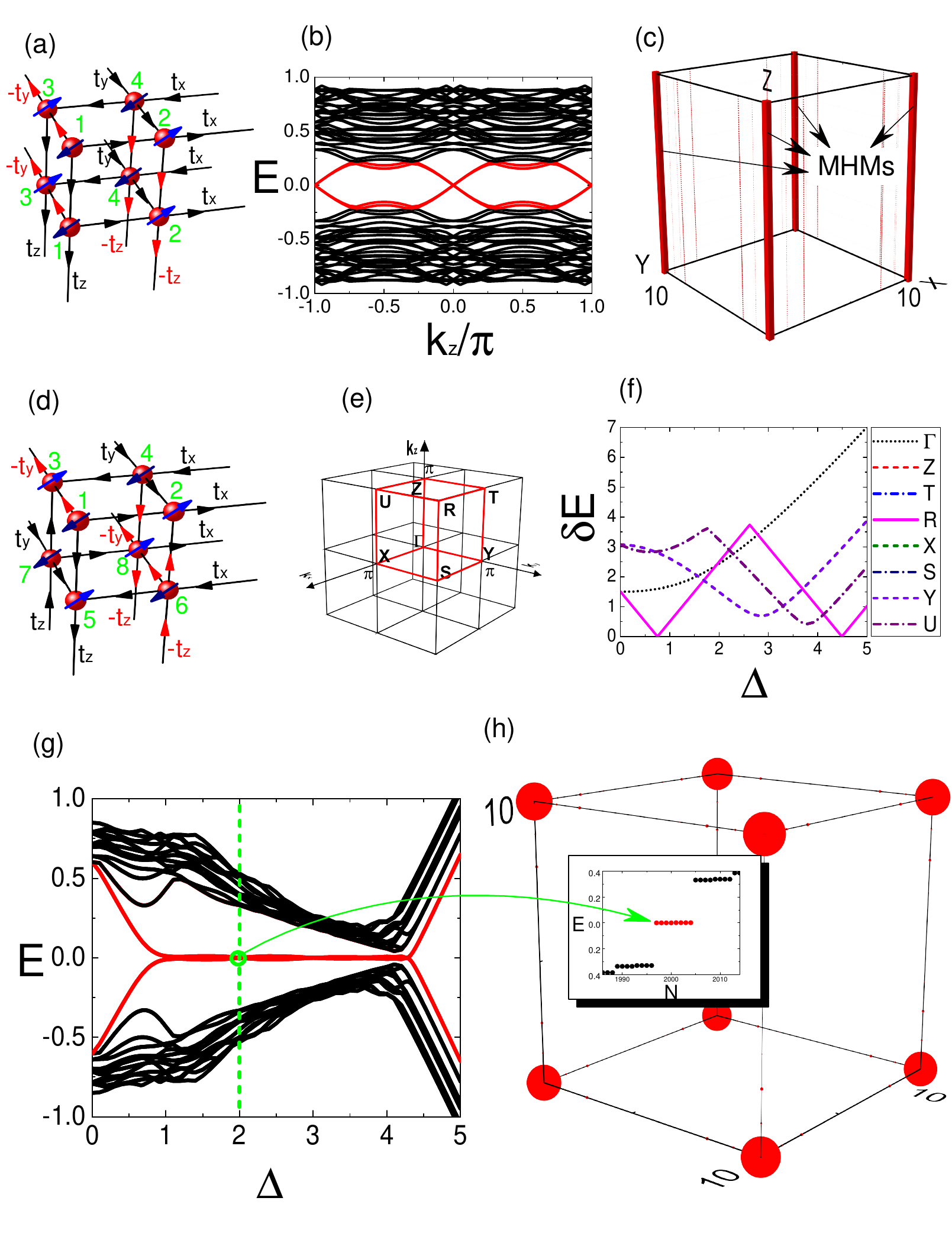}
\caption{(a)(d) The two stacking ways that yield 3D SOTSC and third-order TSCs, respectively. The arrows on the bonds indicate the hopping direction with positive phases, while the colors represent the signs of hopping amplitudes. (b) The energy dispersion of the structure in panel (a) with OBCs applied in the $x$- and $y$-directions and PBCs in the $z$-direction. (c) The density distributions of the MHMs with two hinge modes along each hinge. (e) Highlights the HSPs in a 3D cubic BZ. (f) The gaps at the HSPs of the structure depicted in panel (d) as a function of $\Delta$. The energies at $X,Y,Z$ ($U,S,T$) are threefold degenerate. (g) The low energy spectrum with OBCs imposed on all three directions. (h) The density distributions of the eight MCMs (red dots in inset). The parameters are set to $V=2.0, JS=0.6, t_x=t_y=1.0, \Delta=2.0$. $(\phi_z, t_z)$ are $(0.5\pi, 0.2)$ in (b)(c) and $(0.25\pi, 1.0)$ in (f)-(h).
}
\label{Fig4}
\end{figure}

By stacking the 2D model with specific coupling mechanism between neighboring layers, we can achieve 3D HOTSCs. First, we introduce the coupling way to obtain a 3D SOTSC. In this case, each layer is identical while the interlayer hopping parameters are non-uniform, as illustrated in Fig. \ref{Fig4}(a). The Hamiltonian in $k$-space can be expressed as
\begin{eqnarray}\label{eq9}
H_{hinge}(\mathbf{k})=H(\mathbf{k})+2\xi_{z}\cos k_z s_z\tau_z+2\eta_z\sin k_z s_z,
\end{eqnarray}
with $\xi_{z}=t_z\cos \phi_z$ and $\eta_{z}=t_z\sin \phi_z$. $t_z$ is the hopping amplitude in the $z$-direction, while $\phi_z$ is a uniform phase factor. When $k_z\neq 0$ or $\pi$, the interlayer coupling term $2\eta_z\sin k_z s_z$ destroy the particle-hole symmetry (with the operator given by $\tau_y\sigma_yK$, where $K$ denotes the complex conjugate operator), thus eliminating the zero modes, as shown in Fig. \ref{Fig4}(b). Zero modes persist at $k_z=0$ or $\pi$ in the form of MHMs along each hinge. The density distributions of the MHMs are shown in Fig. \ref{Fig4}(c), where each hinge accommodates a pair of MHMs.

Next, we stack the 2D TSC in a different way to obtain a third-order TSC, as illustrated in Fig. \ref{Fig4}(d). In contrast to the hinge model, the two neighboring layers are relatively rotated by 180 degrees, together with reversing the phase factors of two interlayer hopping bonds, resulting in opposite signs of the hopping phases between any two  parallel neighbor bonds within a unit cell. Consequently, the number of atoms in each unit cell is doubled, and the Hamiltonian can be expressed as follows,
\begin{eqnarray}\label{eq10}
H_{3D}(\mathbf{k})=H'_{x}+H'_{y}+H'_{z}+H'_{on},
\end{eqnarray}
with
\begin{eqnarray}\label{eq11}
H'_{x}&=&\xi_{x}[(1+\cos k_{x})s_{x}\tau_z-\sin k_{x}s_{y}\tau_z]\nonumber\\
&+&\eta_{x}[(1-\cos k_{x})\gamma_zs_{y}\Gamma_z-\sin k_{x}\gamma_zs_{x}\Gamma_z],\nonumber\\
H'_{y}&=&\xi_{y}[(1+\cos k_{y})\gamma_zs_{z}\Gamma_x\tau_z-\sin k_{y}\gamma_zs_{z}\Gamma_y\tau_z]\nonumber\\
&+&\eta_{y}[(1-\cos k_{y})\Gamma_y-\sin k_{y}\Gamma_x],\nonumber\\
H'_z&=&\xi_{z}[(1+\cos k_z)\gamma_x s_z\tau_z-\sin k_z\gamma_y s_z\tau_z]\nonumber\\
&+&\eta_z[(1-\cos k_z)\gamma_y\Gamma_z-\sin k_z\gamma_x\Gamma_z],\nonumber\\
H'_{on}&=&\Delta\tau_x+(JS_{x}\sigma_x+JS_{y}\sigma_y)\gamma_zs_{z}\Gamma_z+V\sigma_z.
\end{eqnarray}
Here $\gamma_{i}$ represent the pauli matrices acting on the sublayer space. It is worth noting that when considering any two of $H'_{x(y,z)}$, the system corresponds to the uncoupled two layers of 2D SOTSCs described by Eq. (\ref{eq2}). We present the gaps of Hamiltonian (\ref{eq11}) as functions of $\Delta$ at the eight HSPs in the 3D BZ in Fig. \ref{Fig4}(f). The corresponding spectra, numerically obtained by imposing OBCs along all three directions, are displayed in Fig. \ref{Fig4}(g). Notably, the gap-closing phenomenon occurs exclusively at point $R$, accompanied by the emergence and disappearance of zero-energy modes. Similar to the two-dimensional case, the eigenvalues at point $R$ can also be analytically derived and expressed as
$\epsilon(R)=\pm\Delta\pm\delta'_{\pm}$, where $\delta'_{\pm}=\sqrt{(2\sqrt{\eta_{x}^{2}+\eta_{y}^{2}+\eta_{z}^{2}}\pm V)^{2}+J^{2}S^{2}}$. These eigenvalues resemble those in Eq. (\ref{eq4}), differing only by the substitution of $\delta$ with $\delta'$. The appearance of the zero-energy modes can also be explained by edge Hamiltonians. We expand the Hamiltonian around point $R$ to second order while applying OBCs in two directions. The Hamiltonian can also be decomposed into two parts, $H_{0}$ and $H_{p}$. the term $H_{p}$ is modified by an additional mass term $2\eta_z\gamma_y\Gamma_z$, compared to Eq. (\ref{eq7}), arising from the interlayer coupling $H'_z$. Following the same procedure as outlined in edge theory, each pair of edge Hamiltonians connected to the same corner exhibits a gap of $2|(\Delta-\delta'_{-})|$ but with different signs from the perspective of any direction (see the Appendix \ref{C1} for details). Consequently, the cubic open boundary system features one Majorana zero mode localized at each of the eight corners, as illustrated in Fig. \ref{Fig4}(h).

\section{Conclusion and discussion}
We have investigated higher-order topological phases based on a conventional $s$-wave superconductor that lacks spin-orbit interaction and complex magnetic orders, but includes a magnetic flux, antiferromagnetic order, and a perpendicular Zeeman field. We have obtained second-order topological superconductivity in two dimensional and three dimensional situations, and third-order topological superconductivity in three dimensional situation.

The experimental realization employs a superconducting substrate patterned with periodically arranged holes (as shown in Fig. \ref{Fig1}), or alternatively, square arrays of superconducting nanowires. Magnetic atoms are positioned atom-by-atom using STM, coupled via itinerant electrons within the superconductor substrate. Theoretical analyses suggest that antiferromagnetic order is achievable by tuning the chemical potential\cite{75,5}. In experiment, antiferromagnetic order has been observed in Mn dimers\cite{27} and chains\cite{16}, and in two-dimensional Cr adatom Shiba lattices\cite{17}. In our model, the magnetic flux and Zeeman field can be applied independently. Thus, the Zeeman field orientation is not restricted to the $z$-axis but can be arbitrarily oriented, provided it remains perpendicular to the moments of the antiferromagnetic order. Previous studies have shown that an applied magnetic field cants the antiferromagnetic moments, inducing magnetization parallel to the field and effectively enhancing the Zeeman field, while maintaining the antiferromagnetic component perpendicular to the Zeeman field\cite{18,6}. Superconducting loops incorporating $\pi$-Josephson junctions can spontaneously generate magnetic flux\cite{74}. Arrays of such junctions, similar to those in Ref. \cite{77}, can be fabricated to produce the required staggered magnetic flux pattern through the designated holes (see Fig. \ref{Fig1}).

The experiment presents two main challenges beyond the need for high-quality materials. First, the limited range of the YSR states extending or RKKY interaction between magnetic atoms, requiring miniaturization the unit cell size of the $\pi$-Josephson junction array to introduce magnetic flux. Experiments have demonstrated that the YSR state of a single magnetic atom can extend to 30 nm\cite{79}. Utilizing quantum dots, this range can be extended to 200 nm\cite{80}. Furthermore, the range of the RKKY interaction is comparable to the superconducting coherence length, reaching nearly micrometer scales\cite{81}. Although most superconducting Josephson loops currently achieve micrometer-scale dimensions\cite{77,82}, $\pi$-Josephson junctions are now fabricated at the atomic scale ($\leq$5 nm)\cite{72,73}. The fabrication of superconducting square loops on the scale of hundreds of nanometers\cite{83} suggests that smaller $\pi$-junction arrays should be feasible with further technological advancements. Second, material selection must ensure that magnetic atom 1 exhibits an anisotropy YSR state, resulting in opposite signs for the hopping amplitude along the $x$ and $y$ axes. Many orbitally anisotropic YSR states have been experimentally observed\cite{24,25,85,87,88,27}. We propose using a $d_{x^{2}-y^{2}}$ state for atom 1, and isotropic states along the $x$ and $y$ directions, such as $d_{z^{2}}$ states, for the remaining three atoms.

Three-dimensional theoretical models of higher-order topological superconductors can be constructed by stacking two-dimensional counterparts. However, experimental realization poses significant challenges. In 3D hinge model (\ref{eq9}), for example, although the hopping phase factor along $z$-direction can be effectively induced by various methods, such as a superconducting current\cite{3,71}, the differing signs of hopping amplitude necessitate a precisely engineered spatial distribution of interlayer magnetic atomic YSR states. This presents a substantial experimental obstacle. Creating the 3D third-order TSC model (\ref{eq10}) experimentally requires even more sophisticated design and is considerably more complex. Despite these challenges, 3D models remain valuable because, in conjunction with the 2D model, they demonstrate an alternative pathway to achieving higher-order topological superconductors, progressing from second- to third-order and from 2D to 3D systems, independent of spin-orbit coupling.

\section{Acknowledgment}

This work was supported by the National Natural Science Foundation of China (Grant Nos. 12375022, 12065011, 11975110, 12204411 and 12075205), Jiangxi Provincial Natural Science Foundation (Grant No. 20252BAC240167), the General Program of Natural Science of Jinggangshan University (Grant No. JZ2403), the Zhejiang Sci-Tech University Scientific Research Start-up Fund (Grant No. 20062318-Y),  and the PhD Starting Fund Program of Tongren University (Grant No. trxyDH2223).

$Data$ $availability$. The data that support the findings of this article are not publicly available. The data are available from the authors upon reasonable request.

\newpage \clearpage
\onecolumngrid
\appendix

\section{Eigenvaules at the four high-symmetric points of the 2D Brillouin zone}\label{A1}

The eigenvalues of Hamiltonian (\ref{eq3}) in the main text at the four high-symmetric points $\Gamma(0,0)$, $X(\pi,0)$, $Y(0,\pi)$, and $M(\pi,\pi)$ can be expressed as follows:

\begin{eqnarray}\label{S1}
\epsilon(\Gamma)&=&\pm\sqrt{\sqrt{A^{2}+4[J^{2}S^{2}(\xi_{x}^2+\xi_{y}^2)+B^{2}]}\pm2B},\nonumber\\
\epsilon(M)&=&\pm\Delta\pm\sqrt{(\sqrt{4\eta_{x}^{2}+4\eta_{y}^{2}}\pm V)^{2}+J^{2}S^{2}},
\end{eqnarray}
where $A=\xi_{x}^2+\xi_{y}^2+\Delta^{2}-J^2S^2-V^2$, $B=\sqrt{V^2(\xi_{x}^2+\xi_{y}^2+\Delta^2)+J^2S^2\Delta^2}$.
\begin{eqnarray}\label{S2}
\epsilon_{1}(X)=\pm\sqrt{C\pm\sqrt{C^2-D^2}},\nonumber\\
\epsilon_{2}(X)=\pm\sqrt{C'\pm\sqrt{C'^2-D'^2}},
\end{eqnarray}
where $C=(2\eta_{x}-V)^2+4\xi_{y}^{2}+\Delta^2+J^2S^2$, $D=\sqrt{[(2\eta_{x}-V)^2-4\xi_{y}^{2}-\Delta^2+J^2S^2]^2+16\xi_{y}^{2}J^2S^2}$, $C'=(2\eta_{x}+V)^2+4\xi_{y}^{2}+\Delta^2+J^2S^2$ and $D'=\sqrt{[(2\eta_{x}+V)^2-4\xi_{y}^{2}-\Delta^2+J^2S^2]^2+16\xi_{y}^{2}J^2S^2}$.

\begin{eqnarray}\label{S3}
\epsilon_{1}(Y)=\pm\sqrt{G\pm\sqrt{G^2-O^2}},\nonumber\\
\epsilon_{2}(Y)=\pm\sqrt{G'\pm\sqrt{G'^2-O'^2}},
\end{eqnarray}
with $G=(2\xi_{x}-V)^2+4\eta_{y}^{2}+\Delta^2+J^2S^2$, $O=\sqrt{[(2\xi_{x}-V)^2-4\eta_{y}^{2}-\Delta^2+J^2S^2]^2+16\eta_{y}^{2}J^2S^2}$, $G'=(2\xi_{x}+V)^2+4\eta_{y}^{2}+\Delta^2+J^2S^2$ and $O'=\sqrt{[(2\xi_{x}+V)^2-4\eta_{y}^{2}-\Delta^2+J^2S^2]^2+16\eta_{y}^{2}J^2S^2}$.

When $\phi=\pi/4$ and $t_{x}=t_{y}$, $\xi_{x}=\eta_{x}=\xi_{y}=\eta_{y}$, $\epsilon(X)=\epsilon(Y)$.

Here $B,C,C',D,D',G,G',O,O'\geq 0$. The gaps at points $\Gamma$, $X$ and $Y$ remain open as long as $JS\neq 0$ and the phase $0<\phi<\pi/2$. However, the gap at point $M$ closes when $\Delta=\sqrt{(2\sqrt{\eta_x^2+\eta_{y}^2}\pm V)^2+J^2S^2}$.

\section{Derivation of the Edge theory in the 2D system}\label{B1}

\subsection{Edge-I and -III}

We begin by expanding the original Hamiltonian $H(\mathbf{k})$ to second order around the point $M(\pi,\pi)$.
\begin{eqnarray}\label{S4}
H_{q_{x},q_{y}}&=&\frac{1}{2}(\xi_{x}s_{x}\tau_z-\eta_{x} s_{y}\Gamma_z)q_{x}^2+(\eta_{x}s_{x}\Gamma_{z}+\xi_{x}s_{y}\tau_z)q_{x}+\frac{1}{2}(\xi_{y}s_{z}\Gamma_x\tau_z-\eta_{y} \Gamma_y)q_{y}^2+(\eta_{y}\Gamma_{x}+\xi_{y}s_{z}\Gamma_y\tau_z)q_{y}\nonumber\\
&+&2\eta_{x}s_{y}\Gamma_z+2\eta_{y}\Gamma_y+V\sigma_z+\Delta\tau_x+(JS_{x}\sigma_x+JS_{y}\sigma_y)s_{z}\Gamma_z.
\end{eqnarray}
From the perspective of edges-I and -III, we apply periodic boundary conditions in the $x$-direction and open boundary conditions in the $y$-direction, substituting $q_{y}$ with $-i\partial_{y}$ in our analysis. The Hamiltonian is divided into two components: $H_{q_{x},q_{y}}=H_{0}(-i\partial_{y})+H_{p}$. By neglecting the $q_{x}^2$ term in $H_{0}$, we obatin
\begin{eqnarray}\label{S5}
H_{0}&=&-\frac{1}{2}(\xi_{y}s_{z}\Gamma_x\tau_z-\eta_{y} \Gamma_y)\partial_{y}^2-i(\eta_{y}\Gamma_{x}+\xi_{y}s_{z}\Gamma_y\tau_z)\partial_{y},\nonumber\\
H_{p}&=&(\eta_{x}s_{x}\Gamma_{z}+\xi_{x}s_{y}\tau_z)q_{x}+2\eta_{y}\Gamma_y+2\eta_{x}s_{y}\Gamma_z+\Delta\tau_x+JS_{x}\sigma_xs_{z}\Gamma_z+JS_{y}\sigma_ys_{z}\Gamma_z+V\sigma_z.
\end{eqnarray}
Following the analysis of the gap structure at point $M$ discussed in the main text, we find that the gap is opened by the $H_{p}$ term. When the gap is small, the impact of $H_{p}$ on $H_{0}$ is small; therefore we can treat $H_{p}$ as a perturbation. We solve $H_{0}$ exactly and then project $H_{p}$ onto the basis of the eigenstates of $H_{0}$. A trial solution of the form $\psi_{\alpha}=\chi_{\alpha}e^{\lambda_{\alpha}y}$ satisfies the equation $H_{0}\psi_{\alpha}=0$. 
The secular equation corresponding to $H_{0}\psi_{\alpha}=0$ is given by $\mathrm{Det}[\frac{1}{2}(\xi_{y}s_{z}\Gamma_x\tau_z-\eta_{y} \Gamma_y)\lambda_{\alpha}^2+i(\eta_{y}\Gamma_{x}+\xi_{y}s_{z}\Gamma_y\tau_z)\lambda_{\alpha}]=0$.
However, we find when neglecting the two $\eta_{y}$ terms does not alter the solution of the equation $H_{0}\psi_{\alpha}=0$. The simplified secular equation is
\begin{eqnarray}\label{S6}
\mathrm{Det}[\frac{1}{2}\xi_{y}s_{z}\Gamma_x\tau_z\lambda_{\alpha}^2+i\xi_{y}s_{z}\Gamma_y\tau_z\lambda_{\alpha}]=0.
\end{eqnarray}
We get $\lambda_{\alpha}=0,\pm$2. To attain a more precise spatial wave function, it is essential to expand the Hamiltonian to higher orders; however, this will not impact the final results of this paper. For simplicity, in the subsequent sections, we will concentrate exclusively on the second-order expansion. From the perspective of edge-I, let $\psi_{\alpha\leq4}=A_{1}\chi_{\alpha\leq4}(A_{2}e^{-2y}+1)$ and $\psi_{4<\alpha\leq8}=A_{1}\chi_{4<\alpha\leq8}(A_{2}e^{2(y-L_{y})}+1)$, where $A_{1}$ is the normalization index and $A_{2}$ is a arbitrary constant. The spinor component of the eigenstate satisfies $s_{y}\sigma_z\Gamma_{0}\tau_{0}\chi_{\alpha}=+\chi_{\alpha}$. We list the eigenvectors as follows:
\begin{eqnarray}\label{S7}
\chi_{1}=|s_{y}=+1\rangle|\sigma_{z}=+1\rangle|\Gamma_{z}=+1\rangle|\tau_{x}=+1\rangle,\nonumber\\
\chi_{2}=|s_{y}=+1\rangle|\sigma_{z}=+1\rangle|\Gamma_{z}=+1\rangle|\tau_{x}=-1\rangle,\nonumber\\
\chi_{3}=|s_{y}=-1\rangle|\sigma_{z}=-1\rangle|\Gamma_{z}=+1\rangle|\tau_{x}=+1\rangle,\nonumber\\
\chi_{4}=|s_{y}=-1\rangle|\sigma_{z}=-1\rangle|\Gamma_{z}=+1\rangle|\tau_{x}=-1\rangle,\nonumber\\
\chi_{5}=|s_{y}=+1\rangle|\sigma_{z}=+1\rangle|\Gamma_{z}=-1\rangle|\tau_{x}=+1\rangle,\nonumber\\
\chi_{6}=|s_{y}=+1\rangle|\sigma_{z}=+1\rangle|\Gamma_{z}=-1\rangle|\tau_{x}=-1\rangle,\nonumber\\
\chi_{7}=|s_{y}=-1\rangle|\sigma_{z}=-1\rangle|\Gamma_{z}=-1\rangle|\tau_{x}=+1\rangle,\nonumber\\
\chi_{8}=|s_{y}=-1\rangle|\sigma_{z}=-1\rangle|\Gamma_{z}=-1\rangle|\tau_{x}=-1\rangle.
\end{eqnarray}
Using the basis above, the matrix elements of $H_{p}$ can be calculated as
\begin{eqnarray}\label{S8}
H_{I,\alpha\beta}=H_{III,\alpha\beta}=\int_{0}^{\infty} dy\psi_{\alpha}^{*}(y)H_{p}(q_{x},-i\partial_{y})\psi_{\beta}(y).
\end{eqnarray}
We obtain the edge Hamiltonians of edges-I and -III within the subspace defined by the eigenvalue of $s_{y}\sigma_z\Gamma_{0}\tau_{0}$ equal +1,
\begin{eqnarray}\label{S8_2}
H_I=H_{III}=\xi_{x}\tau_z\sigma_xq_{x}+2\eta_{x}\Gamma_z\tau_z+2\eta_y\Gamma_y+\Delta\sigma_z
-(JS_{x}\tau_x+JS_{y}\tau_y)\Gamma_z+V\tau_z.
\end{eqnarray}

From the perspective of edge-III, let $\psi_{8<\alpha\leq12}=A_{1}\chi_{8<\alpha\leq12}(A_{2}e^{-2(y+L_{y})}+1)$ and $\psi_{\alpha>12}=A_{1}\chi_{\alpha>12}(A_{2}e^{2y}+1)$. The spinor parts are the eigenvectors with the eigenvalues of $s_{y}\sigma_z\Gamma_{0}\tau_{0}$ equal to -1.
\begin{eqnarray}\label{S9}
\chi_{9}=|s_{y}=-1\rangle|\sigma_{z}=+1\rangle|\Gamma_{z}=+1\rangle|\tau_{x}=+1\rangle,\nonumber\\
\chi_{10}=|s_{y}=-1\rangle|\sigma_{z}=+1\rangle|\Gamma_{z}=+1\rangle|\tau_{x}=-1\rangle,\nonumber\\
\chi_{11}=|s_{y}=+1\rangle|\sigma_{z}=-1\rangle|\Gamma_{z}=+1\rangle|\tau_{x}=+1\rangle,\nonumber\\
\chi_{12}=|s_{y}=+1\rangle|\sigma_{z}=-1\rangle|\Gamma_{z}=+1\rangle|\tau_{x}=-1\rangle,\nonumber\\
\chi_{13}=|s_{y}=-1\rangle|\sigma_{z}=+1\rangle|\Gamma_{z}=-1\rangle|\tau_{x}=+1\rangle,\nonumber\\
\chi_{14}=|s_{y}=-1\rangle|\sigma_{z}=+1\rangle|\Gamma_{z}=-1\rangle|\tau_{x}=-1\rangle,\nonumber\\
\chi_{15}=|s_{y}=+1\rangle|\sigma_{z}=-1\rangle|\Gamma_{z}=-1\rangle|\tau_{x}=+1\rangle,\nonumber\\
\chi_{16}=|s_{y}=+1\rangle|\sigma_{z}=-1\rangle|\Gamma_{z}=-1\rangle|\tau_{x}=-1\rangle.
\end{eqnarray}
The matrix elements of $H_{p}$ can be calculated as
\begin{eqnarray}\label{S10}
H_{III,\alpha\beta}=H_{I,\alpha\beta}=\int_{-\infty}^{0} dy\psi_{\alpha}^{*}(y)H_{p}(q_{x},-i\partial_{y})\psi_{\beta}(y).
\end{eqnarray}
We obtain the edge Hamiltonians of edges-I and -III within the subspace defined by the eigenvalue of $s_{y}\sigma_z\Gamma_{0}\tau_{0}$ equal -1,
\begin{eqnarray}\label{S11}
H_{I}=H_{III}=-\xi_{x}\tau_z\sigma_xq_{x}-2\eta_{x}\Gamma_z\tau_z+2\eta_y\Gamma_y+\Delta\sigma_z
-(JS_{x}\tau_x+JS_{y}\tau_y)\Gamma_z+V\tau_z.
\end{eqnarray}
Therefore, the complete edge Hamitonians of edges-I and -III are
\begin{eqnarray}\label{S11_2}
H_I&=&\xi_{x}s_z\tau_z\sigma_xq_{x}+2\eta_{x}s_z\Gamma_z\tau_z+2\eta_y\Gamma_y+\Delta\sigma_z
-(JS_{x}\tau_x+JS_{y}\tau_y)\Gamma_z+V\tau_z,\nonumber\\
H_{III}&=&-\xi_{x}s_z\tau_z\sigma_xq_{x}-2\eta_{x}s_z\Gamma_z\tau_z+2\eta_y\Gamma_y+\Delta\sigma_z
-(JS_{x}\tau_x+JS_{y}\tau_y)\Gamma_z+V\tau_z,
\end{eqnarray}
with the Pauli matrices $s_z$, $\Gamma_{i}$, $\tau_{i}$ and $\sigma_{i}$ are defined within the space of the sixteen zero-energy states $\psi_{\alpha}$.
\subsection{Edge-II and -IV}
For edges-II and -IV, we apply periodic boundary conditions along the $y$-direction and open boundary conditions along the $x$-direction. The original Hamiltonian can be divided into two parts as $H_{q_{x},q_{y}}=H_{0}(-i\partial_{y})+H_{p}$. By neglecting the $q_{y}^2$ term in $H_{0}$, we obatin
\begin{eqnarray}\label{S12}
H_{0}&=&-\frac{1}{2}(\xi_{x}s_{x}\tau_z-\eta_{x} s_{y}\Gamma_z)\partial_{x}^2-i(\eta_{x}s_{x}\Gamma_{z}+\xi_{x}s_{y}\tau_z)\partial_{x},\nonumber\\
H_{p}&=&(\eta_{y}\Gamma_{x}+\xi_{y}s_{z}\Gamma_y\tau_z)q_{y}+2\eta_{y}\Gamma_y+2\eta_{x}s_{y}\Gamma_z+\Delta\tau_x+JS_{x}\sigma_xs_{z}\Gamma_z+JS_{y}\sigma_ys_{z}\Gamma_z+V\sigma_z.
\end{eqnarray}
We solve $H_{0}$ exactly and treat $H_{p}$ as a perturbation. By using a trial solution $\psi_{\alpha}=\chi_{\alpha}e^{\lambda_{\alpha}x}$, the secular equation of equation $H_{0}\psi_{\alpha}=0$ is given by $\mathrm{Det}[\frac{1}{2}(\xi_{x}s_{x}\tau_z-\eta_{x} s_{y}\Gamma_z)\lambda_{\alpha}^2+i(\eta_{x}s_{x}\Gamma_{z}+\xi_{x}s_{y}\tau_z)\lambda_{\alpha}]=0$.
However, we find when neglecting the two $\eta_{x}$ terms does not alter the solution of the equation $H_{0}\psi_{\alpha}=0$. The simplified secular equation is
\begin{eqnarray}\label{S13}
\mathrm{Det}[\frac{1}{2}\xi_{x}s_{x}\tau_z\lambda_{\alpha}^2+i\xi_{x}s_{y}\tau_z\lambda_{\alpha}]=0.
\end{eqnarray}
We get $\lambda_{\alpha}=\pm$2,0. From the  perspective of edge-II, let $\psi_{\alpha\leq4}=A_{1}\chi_{\alpha\leq4}(A_{2}e^{-2(x+L_{x})}+1)$, $\psi_{4<\alpha\leq8}=A_{1}\chi_{4<\alpha\leq8}(A_{2}e^{2x}+1)$. The spinor component of the eigenstate $\chi_{\alpha}$ satisfies $\Gamma_{y}s_{z}\sigma_z\tau_{0}\chi_{\alpha}=+\chi_{\alpha}$. We list the eigenvectors as follow,
\begin{eqnarray}\label{S14}
\chi_{1}=|\Gamma_{y}=+1\rangle|s_{z}=+1\rangle|\sigma_{z}=+1\rangle|\tau_{x}=-1\rangle,\nonumber\\
\chi_{2}=|\Gamma_{y}=+1\rangle|s_{z}=+1\rangle|\sigma_{z}=+1\rangle|\tau_{x}=+1\rangle,\nonumber\\
\chi_{3}=|\Gamma_{y}=-1\rangle|s_{z}=+1\rangle|\sigma_{z}=-1\rangle|\tau_{x}=-1\rangle,\nonumber\\
\chi_{4}=|\Gamma_{y}=-1\rangle|s_{z}=+1\rangle|\sigma_{z}=-1\rangle|\tau_{x}=+1\rangle,\nonumber\\
\chi_{5}=|\Gamma_{y}=-1\rangle|s_{z}=-1\rangle|\sigma_{z}=+1\rangle|\tau_{x}=-1\rangle,\nonumber\\
\chi_{6}=|\Gamma_{y}=-1\rangle|s_{z}=-1\rangle|\sigma_{z}=+1\rangle|\tau_{x}=+1\rangle,\nonumber\\
\chi_{7}=|\Gamma_{y}=+1\rangle|s_{z}=-1\rangle|\sigma_{z}=-1\rangle|\tau_{x}=-1\rangle,\nonumber\\
\chi_{8}=|\Gamma_{y}=+1\rangle|s_{z}=-1\rangle|\sigma_{z}=-1\rangle|\tau_{x}=+1\rangle.
\end{eqnarray}
The matrix elements of $H_{p}$ can be calculated as
\begin{eqnarray}\label{S15}
H_{II,\alpha\beta}=H_{IV,\alpha\beta}=\int_{-\infty}^{0} dx\psi_{\alpha}^{*}(x)H_{p}(-i\partial_{x},q_{y})\psi_{\beta}(x).
\end{eqnarray}
We obtain the edge Hamiltonians of edges-II and -IV within the subspace defined by the eigenvalue of $\Gamma_{y}s_{z}\sigma_z\tau_{0}$ equal +1,
\begin{eqnarray}\label{S16}
H_{II}=H_{IV}=\xi_{y}\tau_z\sigma_xq_{y}+2\eta_{y}\Gamma_z\tau_z-2\eta_x\Gamma_y-\Delta\sigma_z
-(JS_{x}\tau_x+JS_{y}\tau_y)\Gamma_z+V\tau_z.
\end{eqnarray}
From the perspective of edge-IV, following the same procedure, the spinor component $\chi_{\alpha}$ corresponding to the eigenvectors with eigenvalue of $\Gamma_{y}s_{z}\sigma_z\tau_{0}$ equal to -1.
\begin{eqnarray}\label{S17}
\chi_{9}=|\Gamma_{y}=-1\rangle|s_{z}=+1\rangle|\sigma_{z}=+1\rangle|\tau_{x}=-1\rangle,\nonumber\\
\chi_{10}=|\Gamma_{y}=-1\rangle|s_{z}=+1\rangle|\sigma_{z}=+1\rangle|\tau_{x}=+1\rangle,\nonumber\\
\chi_{11}=|\Gamma_{y}=+1\rangle|s_{z}=+1\rangle|\sigma_{z}=-1\rangle|\tau_{x}=-1\rangle,\nonumber\\
\chi_{12}=|\Gamma_{y}=+1\rangle|s_{z}=+1\rangle|\sigma_{z}=-1\rangle|\tau_{x}=+1\rangle,\nonumber\\
\chi_{13}=|\Gamma_{y}=+1\rangle|s_{z}=-1\rangle|\sigma_{z}=+1\rangle|\tau_{x}=-1\rangle,\nonumber\\
\chi_{14}=|\Gamma_{y}=+1\rangle|s_{z}=-1\rangle|\sigma_{z}=+1\rangle|\tau_{x}=+1\rangle,\nonumber\\
\chi_{15}=|\Gamma_{y}=-1\rangle|s_{z}=-1\rangle|\sigma_{z}=-1\rangle|\tau_{x}=-1\rangle,\nonumber\\
\chi_{16}=|\Gamma_{y}=-1\rangle|s_{z}=-1\rangle|\sigma_{z}=-1\rangle|\tau_{x}=+1\rangle.
\end{eqnarray}
We obtain the edge Hamiltonians of edges-II and -IV within the subspace defined by the eigenvalue of $\Gamma_{y}s_{z}\sigma_z\tau_{0}$ equal -1,
\begin{eqnarray}\label{S18}
H_{IV}=H_{II}=-\xi_{y}\tau_z\sigma_xq_{y}-2\eta_{y}\Gamma_z\tau_z-2\eta_x\Gamma_y-\Delta\sigma_z
-(JS_{x}\tau_x+JS_{y}\tau_y)\Gamma_z+V\tau_z.
\end{eqnarray}
Thus, the compete edge Hamiltonians of edges-II and -IV can be expressed as
\begin{eqnarray}\label{S19}
H_{II}&=&\xi_{y}s_z\tau_z\sigma_xq_{y}+2\eta_{y}s_z\Gamma_z\tau_z-2\eta_x\Gamma_y-\Delta\sigma_z
-(JS_{x}\tau_x+JS_{y}\tau_y)\Gamma_z+V\tau_z,\nonumber\\
H_{IV}&=&-\xi_{y}s_z\tau_z\sigma_xq_{y}-2\eta_{y}s_z\Gamma_z\tau_z-2\eta_x\Gamma_y-\Delta\sigma_z
-(JS_{x}\tau_x+JS_{y}\tau_y)\Gamma_z+V\tau_z.
\end{eqnarray}

In summary, the Hamiltonians for all the four edges are
\begin{eqnarray}\label{S20}
H_I&=&\xi_{x}s_z\tau_z\sigma_xq_{x}+\Delta\sigma_z+2\eta_{x}s_z\Gamma_z\tau_z+2\eta_y\Gamma_y
-(JS_{x}\tau_x+JS_{y}\tau_y)\Gamma_z+V\tau_z,\nonumber\\
H_{II}&=&\xi_{y}s_z\tau_z\sigma_xq_{y}-\Delta\sigma_z+2\eta_{y}s_z\Gamma_z\tau_z-2\eta_x\Gamma_y
-(JS_{x}\tau_x+JS_{y}\tau_y)\Gamma_z+V\tau_z,\nonumber\\
H_{III}&=&-\xi_{x}s_z\tau_z\sigma_xq_{x}+\Delta\sigma_z-2\eta_{x}s_z\Gamma_z\tau_z+2\eta_y\Gamma_y
-(JS_{x}\tau_x+JS_{y}\tau_y)\Gamma_z+V\tau_z,\nonumber\\
H_{IV}&=&-\xi_{y}s_z\tau_z\sigma_xq_{y}-\Delta\sigma_z-2\eta_{y}s_z\Gamma_z\tau_z-2\eta_x\Gamma_y
-(JS_{x}\tau_x+JS_{y}\tau_y)\Gamma_z+V\tau_z.
\end{eqnarray}
We remark that here the Pauli matrices $s_z$, $\Gamma_{i}$, $\tau_{i}$ and $\sigma_{i}$ are defined within the space of the sixteen zero-energy states $\psi_{\alpha}$.
The eigenvalues of each edge Hamiltonian at $q_{x(y)}=0$ is identical and given by,
\begin{eqnarray}\label{S21}
\pm\Delta\pm\sqrt{(\sqrt{4\eta_{x}^{2}+4\eta_{y}^{2}}\pm V)^{2}+J^{2}S^{2}}.
\end{eqnarray}
Taking the counterclockwise direction as the positive direction, which affects both $q_{i}$ and the phase factor $\phi$, the edge Hamiltonians can be compactly written as
\begin{eqnarray}\label{S22}
H_l&=&-iA(l)s_z\tau_z\sigma_x\partial_{l}+\Delta(l)\sigma_z+H_{M},\nonumber\\
H_{M}&=&2\eta(l)s_z\Gamma_z\tau_z+2\eta'(l)\Gamma_y-(JS_{x}\tau_x+JS_{y}\tau_y)\Gamma_z+V\tau_z,
\end{eqnarray}
where $l$=\{I$-$IV\}, $A(l)=\{\xi_{x},\xi_{y},\xi_{x},\xi_{y}\}$, $\Delta(l)=\{\Delta,-\Delta,\Delta,-\Delta\}$, $\eta(l)=\{\eta_{x},\eta_{y},\eta_{x},\eta_{y}\}$ and $\eta'(l)=\{\eta_{y},-\eta_{x},\eta_{y},-\eta_{x}\}$.
Although the terms in $\eta'(l)$ exhibit different signs at different edges, they do not infuence the signs of the gaps.
\section{Derivation of the 3D Edge theory}\label{C1}

\begin{figure}[t]
\centering\includegraphics[width=0.48\textwidth]{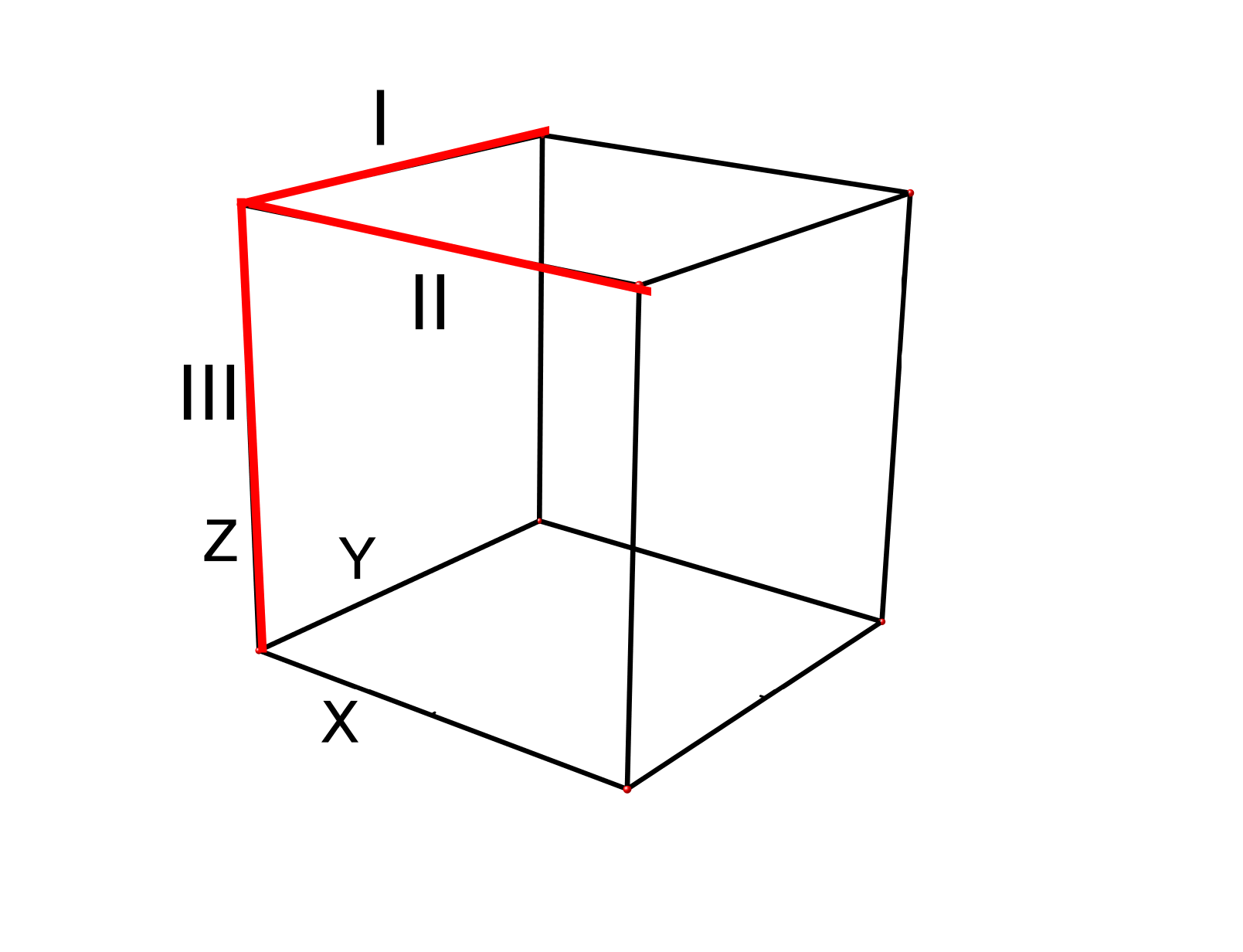}
\caption{The schematic depicts the three-dimensional cubic Brillouin zone. We focus on the three red edges to examine the three-dimensional edge theory.
}
\label{figS1}
\end{figure}

The model is consistent from any direction. In this study, we focus on the three edges connected to a single corner trijunction, as illustrated in Fig. \ref{figS1}. We will present the edge Hamiltonians for these three edges, only taking account of the part calculated from their respective perspectives. This simplification does not affect the overall conclusions. We expand the Hamiltonian $H_{3D}(\mathbf{k})$ around point $R(\pi,\pi,\pi)$ to the second order.
\begin{eqnarray}\label{S23}
H'_{x}&=&\frac{1}{2}(\xi_{x}s_{x}\tau_z-\eta_{x}\gamma_z s_{y}\Gamma_z)q_{x}^2+(\eta_{x}\gamma_zs_{x}\Gamma_{z}+\xi_{x}s_{y}\tau_z)q_{x}+2\eta_{x}\gamma_zs_{y}\Gamma_z,\nonumber\\
H'_{y}&=&\frac{1}{2}(\xi_{y}\gamma_zs_{z}\Gamma_x\tau_z-\eta_{y} \Gamma_y)q_{y}^2+(\eta_{y}\Gamma_{x}+\xi_{y}\gamma_zs_{z}\Gamma_y\tau_z)q_{y}+2\eta_{y}\Gamma_y,\nonumber\\
H'_z&=&\frac{1}{2}(\xi_{z}\gamma_x s_z\tau_z-\eta_z\gamma_y\Gamma_z)q^{2}_{z}+(\eta_z\gamma_x\Gamma_z+\xi_z\gamma_y s_z\tau_z)q_z+2\eta_z\gamma_y\Gamma_z,\nonumber\\
H'_{on}&=&\Delta\tau_x+(JS_{x}\sigma_x+JS_{y}\sigma_y)\gamma_zs_{z}\Gamma_z+V\sigma_z.
\end{eqnarray}
\subsection{From the perspective of $z$-direction}
When focusing on the edge-II, we apply periodic boundary conditions along the $x$-direction and open boundary conditions along the $y$- and $z$- directions. Substituting $q_{y(z)}$ with $-i\partial_{y(z)}$ in the analysis, we divide the Hamiltonian into two parts, expressed as $H'_{q_{x},q_{y},q_{z}}=H'_{0}(-i\partial_{y},-i\partial_{z})+H'_{p}$. By neglecting the $q_{x}^2$ term in $H'_{0}$, we obatin
\begin{eqnarray}\label{S24}
H'_{0}&=&-\frac{1}{2}(\xi_{y}\gamma_zs_{z}\Gamma_x\tau_z-\eta_{y} \Gamma_y)\partial_{y}^2-i(\eta_{y}\Gamma_{x}+\xi_{y}\gamma_zs_{z}\Gamma_y\tau_z)\partial_{y}\nonumber\\
&&-\frac{1}{2}(\xi_{z}\gamma_x s_z\tau_z-\eta_z\gamma_y\Gamma_z)\partial_z^{2}-i(\eta_z\gamma_x\Gamma_z+\xi_z\gamma_y s_z\tau_z)\partial_z,\nonumber\\
H'_{p}&=&(\eta_{x}\gamma_zs_{x}\Gamma_{z}+\xi_{x}s_{y}\tau_z)q_{x}+2\eta_{y}\Gamma_y+2\eta_{x}\gamma_zs_{y}\Gamma_z+2\eta_z\gamma_y\Gamma_z\nonumber\\
&+&\Delta\tau_x+(JS_{x}\sigma_x+JS_{y}\sigma_y)\gamma_zs_{z}\Gamma_z+V\sigma_z.
\end{eqnarray}

For simplicity, we focus exclusively on the spinor components of the wave functions. We choose the spinor component of the eigenstate which satisfies $\gamma_0s _{y}\Gamma_{0}\sigma_z\tau_{0}\chi_{\alpha}=+\chi_{\alpha}$. The components are presented in the form $|\gamma_z\rangle|s_{y}\rangle|\Gamma_{z}\rangle|\sigma_{z}\rangle|\tau_{x}\rangle$,
\begin{eqnarray}\label{S25}
\chi_{1}=|+1\rangle|+1\rangle|+1\rangle|+1\rangle|+1\rangle,
\chi_{2}=|+1\rangle|+1\rangle|+1\rangle|+1\rangle|-1\rangle,\nonumber\\
\chi_{3}=|+1\rangle|+1\rangle|-1\rangle|+1\rangle|+1\rangle,
\chi_{4}=|+1\rangle|+1\rangle|-1\rangle|+1\rangle|-1\rangle,\nonumber\\
\chi_{5}=|+1\rangle|-1\rangle|+1\rangle|-1\rangle|+1\rangle,
\chi_{6}=|+1\rangle|-1\rangle|+1\rangle|-1\rangle|-1\rangle,\nonumber\\
\chi_{7}=|+1\rangle|-1\rangle|-1\rangle|-1\rangle|+1\rangle,
\chi_{8}=|+1\rangle|-1\rangle|-1\rangle|-1\rangle|-1\rangle,\nonumber\\
\chi_{9}=|-1\rangle|+1\rangle|+1\rangle|+1\rangle|+1\rangle,
\chi_{10}=|-1\rangle|+1\rangle|+1\rangle|+1\rangle|-1\rangle,\nonumber\\
\chi_{11}=|-1\rangle|+1\rangle|-1\rangle|+1\rangle|+1\rangle,
\chi_{12}=|-1\rangle|+1\rangle|-1\rangle|+1\rangle|-1\rangle,\nonumber\\
\chi_{13}=|-1\rangle|-1\rangle|+1\rangle|-1\rangle|+1\rangle,
\chi_{14}=|-1\rangle|-1\rangle|+1\rangle|-1\rangle|-1\rangle,\nonumber\\
\chi_{15}=|-1\rangle|-1\rangle|-1\rangle|-1\rangle|+1\rangle,
\chi_{16}=|-1\rangle|-1\rangle|-1\rangle|-1\rangle|-1\rangle.
\end{eqnarray}
We obtain the edge-II part of the edge Hamiltonian as follows:
\begin{eqnarray}\label{S26}
H_{II}&=&\xi_{x}\Gamma_z\sigma_xq_{x}+\Delta\sigma_z+H_{M},\nonumber\\
H_{M}&=&2\eta_{x}\gamma_z\Gamma_z\tau_z+2\eta_y\tau_y+2\eta_z\gamma_y\tau_z
-(JS_{x}\Gamma_x+JS_{y}\Gamma_y)\gamma_z\tau_z+V\Gamma_z.
\end{eqnarray}

For edge-I, we apply periodic boundary conditions in the $y-$direction and open boundary conditions in the $x$- and $z$-directions, substituting $q_{x(z)}$ with $-i\partial_{x(z)}$ in the analysis. We divide the Hamiltonian into two parts with $H'_{q_{x},q_{y},q_{z}}=H'_{0}(-i\partial_{x},-i\partial_{z})+H'_{p}$. By neglecting the $q_{y}^2$ term in $H'_{0}$, we obatin
\begin{eqnarray}\label{S27}
H'_{0}&=&-\frac{1}{2}(\xi_{x}s_{x}\tau_z-\eta_{x}\gamma_z s_{y}\Gamma_z)\partial_{x}^2-i(\eta_{x}\gamma_z s_{x}\Gamma_{z}+\xi_{x}s_{y}\tau_z)\partial_{x}\nonumber\\
&&-\frac{1}{2}(\xi_{z}\gamma_x s_z\tau_z-\eta_z\gamma_y\Gamma_z)\partial_z^{2}-i(\eta_z\gamma_x\Gamma_z+\xi_z\gamma_y s_z\tau_z)\partial_z,\nonumber\\
H'_{p}&=&(\eta_{y}\Gamma_{x}+\xi_{y}\gamma_zs_{z}\Gamma_y\tau_z)q_{y}+2\eta_{y}\Gamma_y+2\eta_{x}\gamma_zs_{y}\Gamma_z+2\eta_z\gamma_y\Gamma_z\nonumber\\
&+&\Delta\tau_x+(JS_{x}\sigma_x+JS_{y}\sigma_y)\gamma_zs_{z}\Gamma_z+V\sigma_z.
\end{eqnarray}

We choose the spinor component of the eigenstate $\chi_{\alpha}$ that satisfies $\gamma_z\Gamma_{y}s_{z}\sigma_z\tau_{0}\chi_{\alpha}=+\chi_{\alpha}$, and present in the form $|\gamma_z\rangle|\Gamma_{y}\rangle|s_{z}\rangle|\sigma_{z}\rangle|\tau_{x}\rangle$.
\begin{eqnarray}\label{S28}
\chi_{1}=|+1\rangle|+1\rangle|+1\rangle|+1\rangle|-1\rangle,
\chi_{2}=|+1\rangle|+1\rangle|+1\rangle|+1\rangle|+1\rangle,\nonumber\\
\chi_{3}=|-1\rangle|-1\rangle|+1\rangle|+1\rangle|-1\rangle,
\chi_{4}=|-1\rangle|-1\rangle|+1\rangle|+1\rangle|+1\rangle,\nonumber\\
\chi_{5}=|+1\rangle|-1\rangle|+1\rangle|-1\rangle|-1\rangle,
\chi_{6}=|+1\rangle|-1\rangle|+1\rangle|-1\rangle|+1\rangle,\nonumber\\
\chi_{7}=|-1\rangle|+1\rangle|+1\rangle|-1\rangle|-1\rangle,
\chi_{8}=|-1\rangle|+1\rangle|+1\rangle|-1\rangle|+1\rangle,\nonumber\\
\chi_{9}=|+1\rangle|-1\rangle|-1\rangle|+1\rangle|-1\rangle,
\chi_{10}=|+1\rangle|-1\rangle|-1\rangle|+1\rangle|+1\rangle,\nonumber\\
\chi_{11}=|-1\rangle|+1\rangle|-1\rangle|+1\rangle|-1\rangle,
\chi_{12}=|-1\rangle|+1\rangle|-1\rangle|+1\rangle|+1\rangle,\nonumber\\
\chi_{13}=|+1\rangle|+1\rangle|-1\rangle|-1\rangle|-1\rangle,
\chi_{14}=|+1\rangle|+1\rangle|-1\rangle|-1\rangle|+1\rangle,\nonumber\\
\chi_{15}=|-1\rangle|-1\rangle|-1\rangle|-1\rangle|-1\rangle,
\chi_{16}=|-1\rangle|-1\rangle|-1\rangle|-1\rangle|+1\rangle.
\end{eqnarray}
The the edge-I part of the edge Hamiltonian can be written as
\begin{eqnarray}\label{S29}
H_{I}&=&\xi_{y}\Gamma_z\sigma_xq_{y}-\Delta\sigma_z+H_{M},\nonumber\\
H_{M}&=&2\eta_{y}\gamma_z\Gamma_z\tau_z-2\eta_z\tau_y-2\eta_x\gamma_y\tau_z
-(JS_{x}\Gamma_x+JS_{y}\Gamma_y)\gamma_z\tau_z+V\Gamma_z.
\end{eqnarray}
The eigenvalues of Eq. (\ref{S26}) and Eq. (\ref{S29}) are identical when $q_{x(y)}=0$,
\begin{eqnarray}
\epsilon=\pm\Delta\pm\sqrt{(2\sqrt{\eta_{x}^{2}+\eta_{y}^{2}+\eta_{z}^2}\pm V)^{2}+J^{2}S^{2}}.
\end{eqnarray}
In the topologically nontrivial phase, the gaps of edge-$I$ and -$II$ are $|\Delta-\sqrt{(2\sqrt{\eta_{x}^{2}+\eta_{y}^{2}+\eta_{z}^2}- V)^{2}+J^{2}S^{2}}|$ but exhibit opposite signs.

\subsection{From the perspective of $y$-direction}

When focusing on the edge-$III$, we apply periodic boundary conditions along the $z$-direction and open boundary conditions along the $x$- and $y$- directions. Substituting $q_{x(y)}$ with $-i\partial_{x(y)}$ in the analysis, we divide the Hamiltonian into two parts, expressed as $H'_{q_{x},q_{y},q_{z}}=H'_{0}(-i\partial_{x},-i\partial_{y})+H'_{p}$. By neglecting the $q_{z}^2$ term in $H'_{0}$, we obatin
\begin{eqnarray}\label{S30}
H'_{0}&=&-\frac{1}{2}(\xi_{x}s_{x}\tau_z-\eta_{x}\gamma_z s_{y}\Gamma_z)\partial_{x}^2-i(\eta_{x}\gamma_z s_{x}\Gamma_{z}+\xi_{x}s_{y}\tau_z)\partial_{x}\nonumber\\
&&-\frac{1}{2}(\xi_{y}\gamma_zs_{z}\Gamma_x\tau_z-\eta_{y} \Gamma_y)\partial_{y}^2-i(\eta_{y}\Gamma_{x}+\xi_{y}\gamma_zs_{z}\Gamma_y\tau_z)\partial_{y},\nonumber\\
H'_{p}&=&(\eta_z\gamma_x\Gamma_z+\xi_z\gamma_y s_z\tau_z)q_z+2\eta_{y}\Gamma_y+2\eta_{x}\gamma_zs_{y}\Gamma_z+2\eta_z\gamma_y\Gamma_z\nonumber\\
&+&\Delta\tau_x+(JS_{x}\sigma_x+JS_{y}\sigma_y)\gamma_zs_{z}\Gamma_z+V\sigma_z.
\end{eqnarray}

We choose the spinor component of the eigenstate that satisfies $\Gamma_{0}s_{z}\gamma_y\sigma_z\tau_{0}\chi_{\alpha}=+\chi_{\alpha}$, and express in the form $|\Gamma_{z}\rangle|s_{z}\rangle|\gamma_y\rangle|\sigma_{z}\rangle|\tau_{x}\rangle$,
\begin{eqnarray}\label{S31}
\chi_{1}=|+1\rangle|+1\rangle|+1\rangle|+1\rangle|-1\rangle,
\chi_{2}=|+1\rangle|+1\rangle|+1\rangle|+1\rangle|+1\rangle,\nonumber\\
\chi_{3}=|+1\rangle|-1\rangle|+1\rangle|-1\rangle|-1\rangle,
\chi_{4}=|+1\rangle|-1\rangle|+1\rangle|-1\rangle|+1\rangle,\nonumber\\
\chi_{5}=|+1\rangle|-1\rangle|-1\rangle|+1\rangle|-1\rangle,
\chi_{6}=|+1\rangle|-1\rangle|-1\rangle|+1\rangle|+1\rangle,\nonumber\\
\chi_{7}=|+1\rangle|+1\rangle|-1\rangle|-1\rangle|-1\rangle,
\chi_{8}=|+1\rangle|+1\rangle|-1\rangle|-1\rangle|+1\rangle,\nonumber\\
\chi_{9}=|-1\rangle|+1\rangle|+1\rangle|+1\rangle|-1\rangle,
\chi_{10}=|-1\rangle|+1\rangle|+1\rangle|+1\rangle|+1\rangle,\nonumber\\
\chi_{11}=|-1\rangle|-1\rangle|+1\rangle|-1\rangle|-1\rangle,
\chi_{12}=|-1\rangle|-1\rangle|+1\rangle|-1\rangle|+1\rangle,\nonumber\\
\chi_{13}=|-1\rangle|-1\rangle|-1\rangle|+1\rangle|-1\rangle,
\chi_{14}=|-1\rangle|-1\rangle|-1\rangle|+1\rangle|+1\rangle,\nonumber\\
\chi_{15}=|-1\rangle|+1\rangle|-1\rangle|-1\rangle|-1\rangle,
\chi_{16}=|-1\rangle|+1\rangle|-1\rangle|-1\rangle|+1\rangle.
\end{eqnarray}

The edge-III part of the edge Hamiltonian can be written as
\begin{eqnarray}\label{S32}
H_{III}&=&\xi_{z}\tau_z\sigma_xq_{z}-\Delta\sigma_z+H_{M},\nonumber\\
H_{M}&=&2\eta_z\gamma_z\Gamma_z+2\eta_y\gamma_y-2\eta_{x}\gamma_z\Gamma_y\tau_z
+(JS_{x}\tau_y-JS_{y}\tau_x)\Gamma_y+V\tau_z.
\end{eqnarray}

When focusing on the edge-II, we choose the spinor component of the eigenstate which satisfies $\Gamma_{0}s_{y}\gamma_0\sigma_z\tau_{0}\chi_{\alpha}=+\chi_{\alpha}$, and present in the form $|\Gamma_{z}\rangle|s_{y}\rangle|\gamma_z\rangle|\sigma_{z}\rangle|\tau_{x}\rangle$,
\begin{eqnarray}\label{S33}
\chi_{1}=|+1\rangle|+1\rangle|+1\rangle|+1\rangle|+1\rangle,
\chi_{2}=|+1\rangle|+1\rangle|+1\rangle|+1\rangle|-1\rangle,\nonumber\\
\chi_{3}=|+1\rangle|-1\rangle|-1\rangle|-1\rangle|+1\rangle,
\chi_{4}=|+1\rangle|-1\rangle|-1\rangle|-1\rangle|-1\rangle,\nonumber\\
\chi_{5}=|+1\rangle|+1\rangle|-1\rangle|+1\rangle|+1\rangle,
\chi_{6}=|+1\rangle|+1\rangle|-1\rangle|+1\rangle|+1\rangle,\nonumber\\
\chi_{7}=|+1\rangle|-1\rangle|+1\rangle|-1\rangle|+1\rangle,
\chi_{8}=|+1\rangle|-1\rangle|+1\rangle|-1\rangle|-1\rangle,\nonumber\\
\chi_{9}=|-1\rangle|+1\rangle|+1\rangle|+1\rangle|+1\rangle,
\chi_{10}=|-1\rangle|+1\rangle|+1\rangle|+1\rangle|-1\rangle,\nonumber\\
\chi_{11}=|-1\rangle|-1\rangle|-1\rangle|-1\rangle|+1\rangle,
\chi_{12}=|-1\rangle|-1\rangle|-1\rangle|-1\rangle|-1\rangle,\nonumber\\
\chi_{13}=|-1\rangle|+1\rangle|-1\rangle|+1\rangle|+1\rangle,
\chi_{14}=|-1\rangle|+1\rangle|-1\rangle|+1\rangle|+1\rangle,\nonumber\\
\chi_{15}=|-1\rangle|-1\rangle|+1\rangle|-1\rangle|+1\rangle,
\chi_{16}=|-1\rangle|-1\rangle|+1\rangle|-1\rangle|-1\rangle.
\end{eqnarray}
The edge-II component of the edge Hamiltonian can be written as
\begin{eqnarray}\label{S34}
H_{II}&=&\xi_{x}\tau_z\sigma_xq_{x}+\Delta\sigma_z+H_{M},\nonumber\\
H_{M}&=&2\eta_{x}\gamma_z\Gamma_z+2\eta_y\gamma_y+2\eta_z\gamma_z\Gamma_y\tau_z
+(JS_{x}\tau_y-JS_{y}\tau_x)\Gamma_y+V\tau_z.
\end{eqnarray}
The eigenvalues of Eq. (\ref{S32}) and Eq. (\ref{S34}) are identical when $q_{z(x)}=0$,
\begin{eqnarray}
\epsilon=\pm\Delta\pm\sqrt{(2\sqrt{\eta_{x}^{2}+\eta_{y}^{2}+\eta_{z}^2}\pm V)^{2}+J^{2}S^{2}}.
\end{eqnarray}
The gaps of edges-II and III are $|\Delta-\sqrt{(2\sqrt{\eta_{x}^{2}+\eta_{y}^{2}+\eta_{z}^2}- V)^{2}+J^{2}S^{2}}|$ but have opposite signs.
\subsection{From the perspective of $x$-direction}

When focusing on the edge-I, we choose the spinor component of the eigenstate which satisfies $s_{z}\Gamma_{y}\gamma_z\sigma_z\tau_{0}\chi_{\alpha}=+\chi_{\alpha}$, and present them in the form $|s_z\rangle|\Gamma_{y}\rangle|\gamma_{z}\rangle|\sigma_{z}\rangle|\tau_{x}\rangle$,
\begin{eqnarray}\label{S35}
\chi_{1}=|+1\rangle|+1\rangle|+1\rangle|+1\rangle|-1\rangle,
\chi_{2}=|+1\rangle|+1\rangle|+1\rangle|+1\rangle|+1\rangle,\nonumber\\
\chi_{3}=|+1\rangle|-1\rangle|-1\rangle|+1\rangle|-1\rangle,
\chi_{4}=|+1\rangle|-1\rangle|-1\rangle|+1\rangle|+1\rangle,\nonumber\\
\chi_{5}=|-1\rangle|-1\rangle|+1\rangle|+1\rangle|-1\rangle,
\chi_{6}=|-1\rangle|-1\rangle|+1\rangle|+1\rangle|+1\rangle,\nonumber\\
\chi_{7}=|-1\rangle|+1\rangle|-1\rangle|+1\rangle|-1\rangle,
\chi_{8}=|-1\rangle|+1\rangle|-1\rangle|+1\rangle|+1\rangle,\nonumber\\
\chi_{9}=|+1\rangle|-1\rangle|+1\rangle|-1\rangle|-1\rangle,
\chi_{10}=|+1\rangle|-1\rangle|+1\rangle|-1\rangle|+1\rangle,\nonumber\\
\chi_{11}=|+1\rangle|+1\rangle|-1\rangle|-1\rangle|-1\rangle,
\chi_{12}=|+1\rangle|+1\rangle|-1\rangle|-1\rangle|+1\rangle,\nonumber\\
\chi_{13}=|-1\rangle|+1\rangle|+1\rangle|-1\rangle|-1\rangle,
\chi_{14}=|-1\rangle|+1\rangle|+1\rangle|-1\rangle|+1\rangle,\nonumber\\
\chi_{15}=|-1\rangle|-1\rangle|-1\rangle|-1\rangle|-1\rangle,
\chi_{16}=|-1\rangle|-1\rangle|-1\rangle|-1\rangle|+1\rangle.
\end{eqnarray}
The edge-I part of the edge Hamiltonians can be written as
\begin{eqnarray}\label{S36}
H_{I}&=&\xi_{y}\gamma_z\sigma_xq_{y}-\Delta\sigma_z+H_{M},\nonumber\\
H_{M}&=&2\eta_{y}\gamma_z\Gamma_z\tau_z-2\eta_x\Gamma_y\tau_z-2\eta_z\tau_y
-(JS_{x}\gamma_x+JS_{y}\gamma_y)\Gamma_z\tau_z+V\gamma_z.
\end{eqnarray}

For edge-III, we select the spinor part of the eigenstate that satisfies $s_{z}\gamma_{y}\Gamma_0\sigma_z\tau_{0}\chi_{\alpha}=+\chi_{\alpha}$, and express them in the form $|s_z\rangle|\gamma_{y}\rangle|\Gamma_{z}\rangle|\sigma_{z}\rangle|\tau_{x}\rangle$,
\begin{eqnarray}\label{S37}
\chi_{1}=|+1\rangle|+1\rangle|+1\rangle|+1\rangle|+1\rangle,
\chi_{2}=|+1\rangle|+1\rangle|+1\rangle|+1\rangle|-1\rangle,\nonumber\\
\chi_{3}=|+1\rangle|+1\rangle|-1\rangle|+1\rangle|+1\rangle,
\chi_{4}=|+1\rangle|+1\rangle|-1\rangle|+1\rangle|-1\rangle,\nonumber\\
\chi_{5}=|-1\rangle|-1\rangle|+1\rangle|+1\rangle|+1\rangle,
\chi_{6}=|-1\rangle|-1\rangle|+1\rangle|+1\rangle|-1\rangle,\nonumber\\
\chi_{7}=|-1\rangle|-1\rangle|-1\rangle|+1\rangle|+1\rangle,
\chi_{8}=|-1\rangle|-1\rangle|-1\rangle|+1\rangle|-1\rangle,\nonumber\\
\chi_{9}=|+1\rangle|-1\rangle|+1\rangle|-1\rangle|+1\rangle,
\chi_{10}=|+1\rangle|-1\rangle|+1\rangle|-1\rangle|-1\rangle,\nonumber\\
\chi_{11}=|+1\rangle|-1\rangle|-1\rangle|-1\rangle|+1\rangle,
\chi_{12}=|+1\rangle|-1\rangle|-1\rangle|-1\rangle|-1\rangle,\nonumber\\
\chi_{13}=|-1\rangle|+1\rangle|+1\rangle|-1\rangle|+1\rangle,
\chi_{14}=|-1\rangle|+1\rangle|+1\rangle|-1\rangle|-1\rangle,\nonumber\\
\chi_{15}=|-1\rangle|+1\rangle|-1\rangle|-1\rangle|+1\rangle,
\chi_{16}=|-1\rangle|+1\rangle|-1\rangle|-1\rangle|-1\rangle.
\end{eqnarray}
The edge-III part of the edge Hamiltonian can be written as
\begin{eqnarray}\label{S38}
H_{III}&=&\xi_{z}\gamma_z\sigma_xq_{z}+\Delta\sigma_z+H_{M},\nonumber\\
H_{M}&=&2\eta_{z}\gamma_z\Gamma_z\tau_z-2\eta_x\Gamma_y\tau_z+2\eta_y\tau_y
-(JS_{x}\gamma_x+JS_{y}\gamma_y)\Gamma_z\tau_z+V\gamma_z.
\end{eqnarray}

The eigenvalues of Eq. (\ref{S36}) and Eq. (\ref{S38}) are also identical when $q_{y(z)}=0$,
\begin{eqnarray}
\epsilon=\pm\Delta\pm\sqrt{(2\sqrt{\eta_{x}^{2}+\eta_{y}^{2}+\eta_{z}^2}\pm V)^{2}+J^{2}S^{2}}.
\end{eqnarray}
The gaps of edges-I and -III still are $|\Delta-\sqrt{(2\sqrt{\eta_{x}^{2}+\eta_{y}^{2}+\eta_{z}^2}- V)^{2}+J^{2}S^{2}}|$ but exhibit opposite signs.

In summary, regardless of the perspective from any direction, the edge Hamiltonians of adjacent edges possess the same eigenvalue when $q_{x(y,z)}=0$,
\begin{eqnarray}
\epsilon(R)=\pm\Delta\pm\sqrt{(2\sqrt{\eta_{x}^{2}+\eta_{y}^{2}+\eta_{z}^2}\pm V)^{2}+J^{2}S^{2}}.
\end{eqnarray}
This is consistent with the gap-closing conditions outlined in the main text.
When the system is topologically nontrivial, the energy gaps are given by $|\Delta-\sqrt{(2\sqrt{\eta_{x}^{2}+\eta_{y}^{2}+\eta_{z}^2}- V)^{2}+J^{2}S^{2}}|$. Assuming that $\sqrt{(2\sqrt{\eta_{x}^{2}+\eta_{y}^{2}+\eta_{z}^2}- V)^{2}+J^{2}S^{2}}<\Delta<\sqrt{(2\sqrt{\eta_{x}^{2}+\eta_{y}^{2}+\eta_{z}^2}+ V)^{2}+J^{2}S^{2}}$, the signs of the gaps correspond to the signs of the terms $\Delta\sigma_z$.

\twocolumngrid


\begin{thebibliography}{99}

\bibitem{Yu1965}L. Yu, Bound state in supercodutors with paramagnetic impurities, Acta Phys. Sin. 21, 75 (1965).
\bibitem{Shiba1968}H. Shiba, Classical spins in superconductors, Progr.14 Theor. Phys. 40, 435 (1968).
\bibitem{Rusinov1969}A. I. Rusinov, Superconductivity near a paramagnetic impurity, Sov. J. Exp. Theor. Phys. Lett 9, 85 (1969).
\bibitem{1}M. M. Vazifeh and M. Franz, Self-organized topological state with Majorana fermions. Phys. Rev. Lett. 111, 206802 (2013).
\bibitem{2}S. Nadj-Perge, I. K. Drozdov, J. Li, H. Chen, S. Jeon, J. Seo, A. H. MacDonald, B. A. Bernevig, and A. Yazdani, Observation
of Majorana fermions in ferromagnetic atomic chains on a superconductor, Science 346, 602 (2014).
\bibitem{3}A. Heimes, P. Kotetes, and G. Schon, Majorana fermions from Shiba states in an antiferromagnetic chain on top of a superconductor, Phys. Rev. B 90, 060507(R) (2014).
\bibitem{4}J. Rontynen and T. Ojanen, Topological Superconductivity and High Chern Numbers in 2D Ferromagnetic Shiba Lattices, Phys. Rev. Lett. 114, 236803 (2015).
\bibitem{75}M. M. Vazifeh and M. Franz, Self-organized topological state with Majorana fermions, Phys. Rev. Lett. 111, 206802 (2013).
\bibitem{5}J. Xiao and J. An, Chiral symmetries and Majorana fermions in coupled magnetic atomic chains on a superconductor, New J. Phys. 17, 113034 (2015).
\bibitem{6}J. Li, T. Neupert, B. A. Bernevig, and A. Yazdani, Manipulating Majorana zero modes on atomic rings with an external magnetic field, Nat. Commun. 7, 10395 (2016).
\bibitem{7} S. Rex, I. V. Gornyi, and A. D. Mirlin, Majorana bound states in magnetic skyrmions imposed onto a superconductor. Phys. Rev. B 100, 064504 (2019)
\bibitem{8}A. Kobia\l{}ka, N. Sedlmayr, and A. Ptok, Majorana bound states in a superconducting Rashba nanowire in the presence of antiferromagnetic order, Phys. Rev. B 103, 125110 (2021).
\bibitem{9} S. A. D\'{\i}az, J. Klinovaja, D. Loss,  and S. Hoffman, Majorana bound states induced by antiferromagnetic skyrmion textures, Phys. Rev. B 104, 214501 (2021).
\bibitem{10}K.-R. Jeon, B. K. Hazra, K. Cho, A. Chakraborty, J.-C. Jeon, H. Han, H. L. Meyerheim, T. Kontos, and S. S. P. Parkin,
Long-range supercurrents through a chiral non-collinear antiferromagnet in lateral Josephson junctions, Nat.Mater. 20, 1358 (2021).
\bibitem{11}L. Schneider, P. Beck, T. Posske, D. Crawford, E. Mascot, S. Rachel, R. Wiesendanger, and J. Wiebe, Topological Shiba bands in artificial spin chains on superconductors, Nat. Phys. 17, 943 (2021).
\bibitem{12}R. Hess, H. F. Legg, D. Loss, and J. Klinovaja, Prevalence of trivial zero-energy subgap states in nonuniform helical spin chains on the surface of superconductors, Phys. Rev. B 106, 104503 (2022).
\bibitem{13} V. Kachin, T. Ojanen, J. L. Lado, and T. Hyart, Effects of electron-electron interactions in the Yu-Shiba-Rusinov lattice model, Phys. Rev. B 107, 174522(2023).
\bibitem{14}P.-X. Shen, V. Perrin, M. Trif, and P. Simon, Majorana-magnon interactions in topological Shiba chains, Phys. Rev. Res. 5, 033207 (2023).
\bibitem{15}D. Crawford, E. Mascot, M. Shimizu, R. Wiesendanger, D. K. Morr, H. O. Jeschke, and S. Rachel, Increased localization of Majorana modes in antiferromagnetic chains on superconductors, Phys. Rev. B 107, 075410 (2023).
\bibitem{16}L. Schneider, P. Beck, L. Rzsa, T. Posske, J. Wiebe, and R. Wiesendanger, Probing the topologically trivial nature of end states in antiferromagnetic atomic chains on superconductors, Nat. Commun. 14, 2742 (2023).
\bibitem{17}M. O. Soldini, F. Kster, G.Wagner, S. Das, A. Aldarawsheh, R. Thomale, S. Lounis, S. S. P. Parkin, P. Sessi, and T. Neupert, Two-dimensional Shiba lattices as a possible platform for crystalline topological superconductivity, Nat. Phys. 19, 1848 (2023).
\bibitem{18}J. Xiao, Q. Hu, and X. Luo, Magnetic flux induced topological superconductivity in magnetic atomic rings, Phys. Rev. B 109, 205420 (2024).
\bibitem{19}L. Cornils, A. Kamlapure, L. Zhou, S. Pradhan, A. A. Khajetoorians, J. Fransson, J. Wiebe, and R. Wiesendanger, Spin-resolved spectroscopy of the Yu-Shiba-Rusinov states of individual atoms, Phys. Rev. Lett. 119, 197002 (2017).
\bibitem{20}H. Kim, A. Palacio-Morales, T. Posske, L. Rzsa, K. Palots, L. Szunyogh, M. Thorwart, and R. Wiesendanger, Toward tailoring Majorana bound states in artificially constructed magnetic atom chains on elemental superconductors, Sci. Adv. 4, eaar5251 (2018).
\bibitem{21}G. Zhang, T. Samuely, N. Iwahara, J. Kamark, C. Wang, P. W. May, J. K. Jochum, O. Onufriienko, P. Szab, S. Zhou, P. Samuely, V. V. Moshchalkov, L. F. Chibotaru, and H.-G. Rubahn, Yu-Shiba-Rusinov bands in ferromagnetic superconducting diamond, Sci. Adv. 6, eaaz2536 (2020).
\bibitem{22}P. Beck, L. Schneider, L. Rzsa, K. Palots, A. Lszlffy, L. Szunyogh, J. Wiebe, and R. Wiesendanger, Spin-orbit coupling induced splitting of Yu-Shiba-Rusinov states in antiferromagnetic dimers, Nat. Commun. 12, 2040 (2021).
\bibitem{23}L. Schneider, P. Beck, J. Neuhaus-Steinmetz, L. Rzsa, T. Posske, J.Wiebe, and R.Wiesendanger, Precursors of Majorana modes and their length-dependent energy oscillations probed at both ends of atomic Shiba chains, Nat. Nanotechnol. 17, 384 (2022).
\bibitem{37}G. C. M\'{e}nard, S. Guissart, C. Brun, S. Pons, V. S. Stolyarov, F. Debontridder, M. V. Leclerc, E. Janod, L. Cario, D. Roditchev, P. Simon, and T. Cren, Coherent long-range magnetic bound states in a superconductor, Nat. Phys. 11, 1013 (2015).
\bibitem{24}M. Ruby, Y. Peng, F. von Oppen, B. W. Heinrich, and K. J. Franke, Orbital Picture of Yu-Shiba-Rusinov Multiplets, Phys. Rev. Lett. 117, 186801 (2016).
\bibitem{85}D.-J. Choi, C. Rubio-Verdu, J. de Bruijckere, M. M. Ugeda, N. Lorente, and J. I. Pascual, Mapping the orbital structure of impurity bound states in a superconductor, Nat Commun 8, 15175 (2017).
\bibitem{25}M. Ruby, B. W. Heinrich, Y. Peng, F. von Oppen, and K. J. Franke, Wave-Function Hybridization in Yu-Shiba-Rusinov Dimers, Phys. Rev. Lett. 120, 156803 (2018).
\bibitem{26}L. Schneider, M. Steinbrecher, L. R\'{o}zsa, J. Bouaziz, K. Palot\'{a}s, M. dos S. Dias, S. Lounis, J. Wiebe, and R. Wiesendanger, Magnetism and in-gap states of 3d transition metal atoms on superconducting Re, npj Quantum Mater. 4, 42 (2019).
\bibitem{32}H. Kim, L. R\'{o}zsa, D. Schreyer, E. Simon and R. Wiesendanger. Long-range focusing of magnetic bound states in superconducting lanthanum. Nat Commun 11, 4573 (2020).
\bibitem{87}F. Kuster, A. M. Montero, F. S. M. Guimaraes, S. Brinker, S. Lounis, S. S. P. Parkin, and P. Sessi, Correlating Josephson supercurrents and Shiba states in quantum spins unconventionally coupled to superconductors. Nat Commun 12, 1108 (2021).
\bibitem{88}F. Kuster, S. Brinker, S. Lounis, S. S. P. Parkin, and P. Sessi, Long range and highly tunable interaction between local spins coupled to a superconducting condensate, Nat Commun 12, 6722 (2021).
\bibitem{27}P. Beck, L. Schneider, L. R\'{o}zsa, K. Palot\'{a}s, A. L\'{a}szl\'{o}ffy, L. Szunyogh, J. Wiebe, and R. Wiesendanger , Spin-orbit coupling induced splitting of Yu-Shiba-Rusinov states in  antiferromagnetic dimers,  Nat Commun 12, 2040 (2021).
\bibitem{28}L. Schneider, P. Beck, T. Posske, D. Crawford, E. Mascot, S. Rachel, R. Wiesendanger, and J. Wiebe, Topological Shiba bands in artificial spin chains on superconductors, Nat. Phys. 17, 943 (2021).
\bibitem{29} H. M\"uller, M. Eckstein, and S. Viola Kusminskiy, Control of Yu-Shiba-Rusinov States through a Bosonic Mode, Phys. Rev. Lett. 130, 106905 (2023).
\bibitem{30} M. Uldemolins, F. Massee, T. Cren, A. Mesaros, and P. Simon, Anisotropy of Yu-Shiba-Rusinov states in $NbSe_{2}$, Phys. Rev. B 110, 224519 (2024).
\bibitem{31}L. M. R\"utten, H. Schmid, E. Liebhaber, G. Franceschi, A. Yazdani, G. Reecht, K. Rossnagel, F. von Oppen, and K. J. Franke, Wave Function Engineering on Superconducting Substrates: Chiral Yu-Shiba-Rusinov Molecules. ACS Nano 18, 30798 (2024).


\bibitem{39}Y. Volpez, D. Loss, and J. Klinovaja, Second-Order Topological Superconductivity in $\pi$-Junction Rashba Layers, Phys. Rev. Lett. 122, 126402 (2019).
\bibitem{40} S. Franca, D. V. Efremov, and I. C. Fulga, Phase-tunable second-order topological superconductor,  Phys. Rev. B 100, 075415 (2019).
\bibitem{41}K. Laubscher, D. Chughtai, D. Loss, and J. Klinovaja, Kramers pairs of Majorana corner states in a topological insulator bilayer, Phys. Rev. B 102, 195401 (2020).
\bibitem{42}R. Nakai and K. Nomura, Higher-order topological superconductor phases in a multilayer system,  Phys. Rev. B 108, 184517 (2023).
\bibitem{43}Z. Yan, F. Song, and Z. Wang, Majorana Corner Modes in a High-Temperature Platform,  Phys. Rev. Lett. 121, 096803 (2018).
\bibitem{44}X. Zhu, Tunable Majorana corner states in a two-dimensional second-order topological superconductor induced by magnetic fields, Phys. Rev. B 97, 205134 (2018).
\bibitem{442}T. Liu, J. J. He, and F. Nori, Majorana corner states in a two-dimensional magnetic topological insulator on a high-temperature superconductor, Phys. Rev. B 98, 245413 (2018).
\bibitem{45}Y. Wang, M. Lin, and T. L. Hughes, Weak-pairing higher order topological superconductors. Phys. Rev. B 98, 165144 (2018).
\bibitem{46}Z. Yan, Higher-Order Topological Odd-Parity Superconductors, Phys. Rev. Lett. 123, 177001 (2019).
\bibitem{47}X. Zhu, Second-Order Topological Superconductors with Mixed Pairing,  Phys. Rev. Lett. 122, 236401 (2019).
\bibitem{48}Y.-T. Hsu, W. S. Cole, R.-X. Zhang, and J. D. Sau, Inversion-protected Higher-order Topological Superconductivity in Monolayer $WTe_{2}$, Phys. Rev. Lett. 125, 097001 (2020).
\bibitem{49}M. Kheirkhah, D. Zhu, J. Maciejko, and Z. Yan, Corner- and sublattice-sensitive Majorana zero modes on the kagome lattice. Phys. Rev. B 106, 085420 (2022).
\bibitem{50}B. Lu and Y. Zhang, Tunable Majorana corner modes by orbital-dependent exchange interaction in a two-dimensional topological superconductor, J. Phys.: Condens. Matter 34, 305302 (2022).
\bibitem{51}T. Li, M. Geier, J. Ingham, and H. D Scammell, Higher-order topological superconductivity from repulsive interactions in kagome and honeycomb systems, 2D Mater. 9, 015031 (2022).
\bibitem{52}R. Ghadimi, S. H. Lee, and B.-J. Yang, Boundary-obstructed topological superconductor in buckled honeycomb lattice under perpendicular electric field, Phys. Rev. B 107, 224511 (2023).
\bibitem{562}R.-X. Zhang, W. S. Cole, X. Wu, and S. Das Sarma, Higher-Order Topology and Nodal Topological Superconductivity in Fe(Se,Te) Heterostructures, Phys. Rev. Lett. 123, 167001 (2019).
\bibitem{53}Y.-X. Li and C.-C. Liu, Majorana corner modes and tunable patterns in an altermagnet heterostructure, Phys. Rev. B 108, 205410 (2023).
\bibitem{54}K. H. Wong, M. R. Hirsbrunner, J. Gliozzi, A. Malik, B. Bradlyn, T. L. Hughes, and D. K. Morr, Higher order topological superconductivity in magnet-superconductor hybrid systems, npj Quantum Mater. 8, 31 (2023).
\bibitem{55}P. Chatterjee, A. K. Ghosh, A. K. Nandy, and A. Saha, Second-order topological superconductor via noncollinear magnetic texture, Phys. Rev. B 109, L041409 (2024).
\bibitem{70}K. H. Wong, J. Gliozzi, M. R. Hirsbrunner, A. Malik, B. Bradlyn, T. L. Hughes, and D. K. Morr, Competing higher order topological superconducting phases in triangular lattice magnet-superconductor hybrid systems, Phys. Rev. B 109, 144521 (2024).
\bibitem{56}X.-H. Pan, K.-J. Yang, L. Chen, G. Xu, C.-X. Liu, and X. Liu, Lattice-Symmetry-Assisted Second-Order Topological Superconductors and Majorana Patterns, Phys. Rev. Lett. 123, 156801 (2019).

\bibitem{57}Y. Peng and Y. Xu, Proximity-induced Majorana hinge modes in antiferromagnetic topological insulators, Phys. Rev. B 99, 195431 (2019).
\bibitem{58}Y.-J. Wu, J. Hou, Y.-M. Li, X.-W. Luo, X. Shi, and C. Zhang, In-Plane Zeeman-Field-Induced Majorana Corner and Hinge Modes in an $s$-Wave Superconductor Heterostructure, Phys. Rev. Lett. 124, 227001 (2020).

\bibitem{59} M. Kheirkhah, Y. Nagai, C. Chen, and F. Marsiglio, Majorana corner flat bands in two-dimensional second-order topological superconductors, Phys. Rev. B 101, 104502 (2020).
\bibitem{60} X.-J. Luo, X.-H. Pan, and X. Liu, Higher-order topological superconductors based on weak topological insulators, Phys. Rev. B 104, 104510 (2021).
\bibitem{61}H. D. Scammell, J. Ingham, M. Geier, and T. Li, Intrinsic first- and higher-order topological superconductivity in a doped topological insulator, Phys. Rev. B 105, 195149 (2022).
\bibitem{62}X.-T. Chen, C.-H. Liu, D.-H. Xu, and C.-Z. Chen, Majorana Corner Modes and Flat-Band Majorana Edge Modes in Superconductor/Topological-Insulator/Superconductor Junctions, Chin. Phys. Lett. 40, 097403(2023).

\bibitem{95}Y. Kim, M. Cheng, B. Bauer, R. M. Lutchyn, and S. Das Sarma, Helical order in one-dimensional magnetic atom chains and possible emergence of Majorana bound
states, Phys. Rev. B 90, 060401(R) (2014).

\bibitem{63}W. A. Benalcazar, B. A. Bernevig, and T. L. Hughes, Electric multipole moments, topological multipole moment pumping, and chiral hinge states in crystalline insulators, Phys. Rev. B 96, 245115 (2017).
\bibitem{64} W. A. Benalcazar, B. A. Bernevig, and T. L. Hughes, Quantized electric multipole insulators, Science 357, 61-66 (2017).
\bibitem{65}B. Kang, K. Shiozaki, and G. Y. Cho, Many-body order parameters for multipoles in solids, Phys. Rev. B 100, 245134 (2019).
\bibitem{66}W. A. Wheeler, L. K. Wagner, and T. L. Hughes, Many-body electric multipole operators in extended systems, Phys. Rev. B 100, 245135 (2019).
\bibitem{67} C.-A. Li, B. Fu, Z.-A. Hu, J. Li, and S.-Q. Shen, Topological Phase Transitions in Disordered Electric Quadrupole Insulators, Phys. Rev. Lett. 125, 166801 (2020).
\bibitem{69}R. Jackiw and C. Rebbi, Solitons with fermion number 1/2, Phys. Rev. D 13, 3398(1976).
\bibitem{89}S. Ono, L. Trifunovic, and H. Watanabe, Difficulties in operator-based formulation of the bulk quadrupole moment, Phys. Rev. B 100, 245133 (2019).

\bibitem{90}A. Agarwala, V. Juricic, and B. Roy, Higher-order topological insulators in amorphous solids, Phys. Rev. Res. 2, 012067 (2020).
\bibitem{91}Y.-B. Yang, K. Li, L.-M. Duan, and Y. Xu, Type-II quadrupole topological insulators, Phys. Rev. Res. 2, 033029 (2020).
\bibitem{92}T. Peng, C.-B. Hua, R. Chen, Z.-R. Liu, H.-M. Huang, and B. Zhou, Density-driven higher-order topological phase transitions in amorphous solids, Phys. Rev. B 106, 125310 (2022).
\bibitem{93}X. Wang, Y.-X. Li, and T. Zhou, Higher order topological state induced by $d$-wave competing orders in high-$\mathbf{T}_c$ superconductor based heterostructure, Phys. Rev. B 106, 205128 (2022).
\bibitem{94}Y.-B. Yang, J.-H. Wang, K. Li and Y. Xu, Higher-order topological phases in crystalline and non-crystalline systems: a review, J. Phys.: Condens. Matter 36, 283002 (2024).
\bibitem{74}L. N. Bulaevski\v{i}, V. V.  Kuzi\v{i}, and  A. A. Sobyanin, Superconducting system with weak coupling to the current in the ground state, JETP Lett. 25, 290 (1977).

\bibitem{77}S. M. Frolov, M. J. A. Stoutimore, T. A. Crane, D. J. Van Harlingen, V. A. Oboznov, V. V. Ryazanov, A. Ruosi, C. Granata, and M. Russo, Imaging spontaneous currents in superconducting arrays of $\pi$-junctions, Nat. Phys. 4, 32 (2008).
\bibitem{79} H. Kim, L. R\'{o}zsa, D. Schreyer, E. Simon, and R. Wiesendanger, Long-range focusing of magnetic bound states in superconducting lanthanum, Nat. Commun. 11, 4573 (2020).
\bibitem{80}Z. Scherubl, G. Fulop, C. P. Moca, J. Gramich, A. Baumgartner, P. Makk, T. Elalaily, C. Schonenberger, J. Nygard, G. Zarand, and S. Csonka, Large spatial extension of the zero-energy Yu-Shiba-Rusinov state in a magnetic field, Nat. Commun. 11, 1834 (2020).
\bibitem{81} J. F. Annett, Superconductivity, Superfluids and Condensates. New York: Oxford university press (2004).
\bibitem{82}L. V. Abdurakhimov, I. Mahboob, H. Toida, K. Kakuyanagi, and S. Saito, A long-lived capacitively shunted flux qubit embedded in a 3D cavity, Appl. Phys. Lett. 115, 262601 (2019).
\bibitem{73}H. G. Ahmad, M. Minutillo, R. Capecelatro, A. Pal, R. Caruso, G. Passarelli, M. G. Blamire, F. Tafuri, P. Lucignano, and D. Massarotti, Coexistence and tuning of spin-singlet and triplet transport in spin-filter Josephson junctions, Commun. Phys. 5, 2 (2022).
\bibitem{72}K. Senapati, M. Blamire, and Z. Barber, Spin-filter Josephson junctions, Nature Mater 10, 849 (2011).
\bibitem{83}Y. Li  and X. Xu  and M.-H. Lee  and M.-W. Chu, and C. L. Chien, Observation of half-quantum flux in the unconventional superconductor $\beta$-Bi$_{2}$Pd, Science 366, 238 (2019).
\bibitem{71}P. Kotetes,Topological superconductivity in Rashba semiconductors without a Zeeman field, Phys. Rev. B 92, 014514 (2015).







\end{thebibliography}
\end{document}